\documentclass[twocolumn,prb,superscriptaddress,floatfix,aps]{revtex4-2}
\pdfoutput=1

\usepackage[utf8]{inputenc}
\usepackage{caption}
\usepackage{subcaption}
\usepackage{CJKutf8}
\usepackage{siunitx}
\usepackage{graphicx}
\usepackage{float}
\usepackage{amsmath}
\usepackage{amssymb}
\usepackage{dcolumn}
\usepackage{bm}
\usepackage{xcolor}
\usepackage{xfrac}
\usepackage[colorlinks=true, allcolors=black]{hyperref}
\usepackage{chemformula}
\usepackage[normalem]{ulem}
\usepackage{multirow}
\DeclareSIUnit\angstrom{\text {Å}}

\def\Rtbm{$R\overline{3}m$}
\def\Psmmc{$P6_3/mmc$}
\def\Psmmm{$P6/mmm$}
\def\ReMoB{\ch{Re_{0.10}Mo_{0.90}B2}}
\def\MoB{\ch{MoB2}}
\def\Tc{$T_\mathrm{c}$}
\newcommand{\TC}{T_{\mathrm{c}}}    
\definecolor{mag}{RGB}{255,0,255}
\def\micro{\ensuremath{\upmu}}
\begin{document}
\begin{CJK*}{UTF8}{gbsn} 

\title{{Superconductivity in pressurized \ReMoB}}

\author{S. Sinha}
\affiliation{Department of Physics, University of Florida, Gainesville, Florida 32611, USA}
\author{J. Lim}
\affiliation{Department of Physics, University of Florida, Gainesville, Florida 32611, USA}
\author{Z. Li}
\affiliation{Department of Physics, University of Florida, Gainesville, Florida 32611, USA}
\author{J. S. Kim}
\affiliation{Department of Physics, University of Florida, Gainesville, Florida 32611, USA}
\author{A. C. Hire}
\affiliation{Department of Materials Science and  Engineering, University of Florida, Gainesville, Florida 32611, USA}
\affiliation{Quantum Theory Project, University of Florida, Gainesville, Florida 32611, USA}
\author{P. M. Dee}
\affiliation{Department of Physics, University of Florida, Gainesville, Florida 32611, USA}
\affiliation{Department of Materials Science and  Engineering, University of Florida, Gainesville, Florida 32611, USA}
\author{R. S. Kumar}
\affiliation{Department of Physics, University of Illinois Chicago, Chicago, Illinois 60607, USA}
\author{D.~Popov}
\affiliation{HPCAT, X-ray Science Division, Argonne National Laboratory, Argonne, Illinois 60439, USA}
\author{R. J. Hemley}
\affiliation{Department of Physics, Chemistry, and Earth and Environmental Sciences, University of Illinois Chicago, Chicago, Illinois 60607, USA}
\author{R. G. Hennig}
\affiliation{Department of Materials Science and  Engineering, University of Florida, Gainesville, Florida 32611, USA}
\affiliation{Quantum Theory Project, University of Florida, Gainesville, Florida 32611, USA}
\author{P. J. Hirschfeld}
\affiliation{Department of Physics, University of Florida, Gainesville, Florida 32611, USA}
\author{G. R. Stewart}
\affiliation{Department of Physics, University of Florida, Gainesville, Florida 32611, USA}
\author{J. J. Hamlin}
\affiliation{Department of Physics, University of Florida, Gainesville, Florida 32611, USA}
\date{\today}

\begin{abstract}  The recent surprising discovery of superconductivity with critical temperature $\TC = \SI{32}{K}$ in \ch{MoB2} above 70 GPa has led to the search for related materials that may superconduct at similarly high \Tc\ values and lower pressures.
We have studied the superconducting and structural properties of \ReMoB\ to \SI{170}{GPa}. 
A structural phase transition from \Rtbm\ to \Psmmm\ commences at \SI{48}{GPa},
with the first signatures of superconductivity appearing above \SI{44}{GPa}.
The critical temperature is observed to increase with pressure.
A complete resistive transition is observed only above \SI{150}{GPa}, where the highest onset \Tc~of \SI{30}{K} is also achieved. 
Upon releasing pressure, the high pressure superconducting phase is found to be metastable.
During unloading, a complete resistive superconducting transition is observed all the way down to \SI{20}{GPa} (with onset $\TC \sim \SI{20}{K}$).
Our results suggest that the \Psmmm\ structure is responsible for the observed superconductivity.

\end{abstract}

\maketitle
\end{CJK*}

\section{Introduction} 
The search for high-temperature superconductors has been a focus of study for several decades.
Although some hydrides have been reported to show very high \Tc, approaching room temperature, the pressure required to achieve this \Tc~is quite high. 
For example, \ch{H$_3$S} is reported to have a \Tc~of \SI{203}{K} at \SI{155}{GPa}~\cite{Drozdov2015}.
At \SI{190}{GPa}, \ch{LaH10} is reported to reach a \Tc~of \SI{260}{K}~\cite{LaH10,Mozaffari2019}.
The extremely high pressures required to produce this behavior renders these compounds 
(at present) impractical for real-world applications.
The very high critical temperatures in the hydride superconductors are believed to be related to a conventional, electron-phonon coupling mechanism, where, under certain circumstances, low atomic masses can contribute to high \Tc\ values~\cite{AnPickett2001,Mazin2003}.
This motivates a search for non-hydride light-element compounds that may exhibit high \Tc\ values at lower pressures.

One such compound is \ch{MgB2}, which shows a remarkably high \Tc~of \SI{39}{K} at ambient pressure.
For two decades following the discovery of superconductivity in \ch{MgB2}, the study of other diborides did not yield materials with similarly high \Tc\ values~\cite{buzea_review_2001,budko_superconductivity_2015,CarvalhodeCastroSene2024}.
However, the recent discovery of superconductivity in \MoB~\cite{MoB2_superconductivity} with a \Tc\ reaching \SI{32}{K} opens new possibilities for superconductivity in transition metal diborides. 
These authors reported that the high-pressure superconducting phase of \MoB\ shares the same \Psmmm\ crystal structure as  \MoB.
A study by Quan et al.~\cite{Quan2021} claims that although \ch{MoB2} is also a electron-phonon coupling superconductor similar to \ch{MgB2}, in case of \MoB, a strong electron-phonon coupling is provided by Mo atoms and B atoms help in further enhancing \Tc\ at high pressure. 
This is unlike \ch{MgB2}, where electron-phonon coupling is due to B atoms and Mg atoms primarily act as electron donors~\cite{BUDKO2015142, Hinks2001}.
The material, \MoB\ undergoes a structural phase transition from \Rtbm\ to \Psmmm\ at $\sim \SI{65}{GPa}$.
Following this discovery, we explored  superconductivity in \ch{WB2}~\cite{Lim_WB2_2021} ($\TC \sim \SI{17}{K}$ at \SI{90}{GPa}) and Nb substituted \MoB\ ($\TC < \SI{8}{K}$)~\cite{Hire_NbxMoxB2_2022,Lim2023}.
Although \ch{WB2} exists in $P6_3/mmc$ structure throughout the measured pressure range, DFT calculations suggest that the superconductivity observed in \ch{WB2} is due to stacking faults which have a local structure corresponding to the \Psmmm\ (\ch{MgB2}) phase.
Niobium substitution (Nb$_{x}$Mo$_{1-x}$B$_2$) stabilizes the \Psmmm\ structure at ambient pressure over the entire range of concentrations studied ($x$ = 0.1, 0.25, 0.5, and 0.75), and produces superconductivity at ambient pressure ($\TC \sim 6 - \SI{8}{K}$)~\cite{Hire_NbxMoxB2_2022}.
However, \Tc\ initially goes down with pressure and does not exceed \SI{8}{K} to pressures of at least \SI{170}{GPa}.
Partial substitution by Zr, Hf, or Ta in \MoB\ also leads to the \Psmmm\ structure and superconductivity at ambient pressure with measured \Tc s in the range of $2.4 - \SI{8.5}{K}$~\cite{Dee2024}.
These studies indicate that the \Psmmm\ structure has a positive correlation with the superconducting nature of diboride materials.
However, this structure, by itself, is not a sufficient condition for either high \Tc\ or even superconductivity.

Having found that Nb substitution lowers the maximum \Tc\ found in \MoB, we turned to Re substitution.
Re is more electron-rich than both Mo and Nb, suggesting the possibility that Re substitution might have the opposite effect on \Tc\ as Nb substitution.
Moreover, Re atoms (atomic radius: \SI{188}{pm}~\cite{clementi1967atomic}) are very similar in size compared to Mo atoms (atomic radius: \SI{190}{pm}) so Re should be able to replace Mo seamlessly.
However, \ch{ReB2} realizes the \Psmmc\ structure~\cite{RheniumDiboride, Jain2013} at ambient pressure while \ch{MoB2} realizes \Rtbm\ structure at ambient pressure~\cite{MoB2_superconductivity}.
We find that substituting Re for Mo at the 10\% level produces a \Tc\ vs pressure phase diagram very similar to that of pure \MoB\, albeit with a slightly suppressed maximum \Tc\.
However, the structural behavior with pressure show significant differences compared to pure \MoB, which provides deeper understanding of the superconducting nature in both these materials.
Notably the superconductivity only begins to appear once the sample begins to transition to the \ch{MgB2}-like \Psmmm\ structure.

\section{Methods}
\label{sec:methods}
A sample of \ReMoB~sample was synthesized via arc melting rhenium, molybdenum and boron in the ratio of approximately 1:4.6:1.2, by molar weight~\cite{arc_melting,Hire_NbxMoxB2_2022}.
A piece of sample with dimensions $\sim$50$\times$50$\times$\SI{10}{\micro \m^3} was placed inside an Almax OmniDAC diamond anvil cell for resistivity measurements.
The electrodes are connected in van der Pauw (VDP) four probe configuration~\cite{VanderPauw}. 
An ideal VDP configuration on an isotropic sample of uniform thickness, with electrodes attached at the very edge of a sample should give similar resistance in both the possible voltage/current electrode configurations.
However, in DACs, deviations from optimal conditions frequently arise due to non-uniform sample thickness and difficulty in placing electrodes close to the edges, leading to errors in resistivity estimation.

In this study, the ratio of resistance in the two different electrodesconfigurations (measured at \SI{18}{GPa}) at room temperature was 1.17.
Errors in resistivity due to non-ideal electrode placement and finite sample thickness were not taken into consideration and are likely not much larger than 30\%~\cite{Wu_VanderPauw}.
Furthermore, measuring change in sample dimensions and electrode dimensions with pressure is difficult.
Hence, sample dimensions at ambient pressure are used for resistivity estimations at high pressure and any change in sample thickness with pressure is ignored.

Tiny pieces of dimensions $\sim$100$\times$100$\times$\SI{10}{\micro\m^3} taken from the large sample were also packed into two symmetric diamond anvil cells for X-ray diffraction (XRD) measurements~\cite{jayaraman1983diamond}.
In the first XRD measurement using \SI{300}{\micro \m} diamond culets, pressure was increased up to \SI{73}{GPa} before lowering back to ambient pressure.
In the second XRD run with \SI{150}{\micro\m} culets, a maximum pressure of \SI{168}{GPa} was achieved. The XRD measurements were conducted at room temperature.
For resistivity measurements, a maximum pressure of \SI{160}{GPa} (at \SI{10}{K}) was achieved before releasing pressure.
A few measurements were also taken when releasing pressure in case of both XRD and resistivity measurements.
In order to match the sample conditions from the electrical resistance measurements, no pressure medium was used for the XRD experiments.

Ruby fluorescence and the Raman spectrum of the diamond culet were used for pressure estimation below and above \SI{100}{GPa}, respectively~\cite{chijioke_ruby_2005, Akahama2006}.
Ruby measurements were done at both \SI{292}{K} and \SI{10}{K}, while Raman spectroscopy was performed only at \SI{292}{K}.
For pressure estimation at \SI{10}{K} at higher pressures, the pressure difference between \SI{292}{K} and \SI{10}{K} was estimated to be $\sim \SI{20}{GPa}$ based off of
ruby measurements performed at lower pressures.
Further details on XRD and resistivity methods can be found in Ref.~\cite{Lim2023}. Rietveld  refinements on high pressure XRD data were carried out using GSAS-II software~\cite{toby_gsas-ii_2013}. 

\begin{figure}
    \begin{subfigure}[b]{0.48\textwidth}
        \includegraphics[width=\columnwidth]{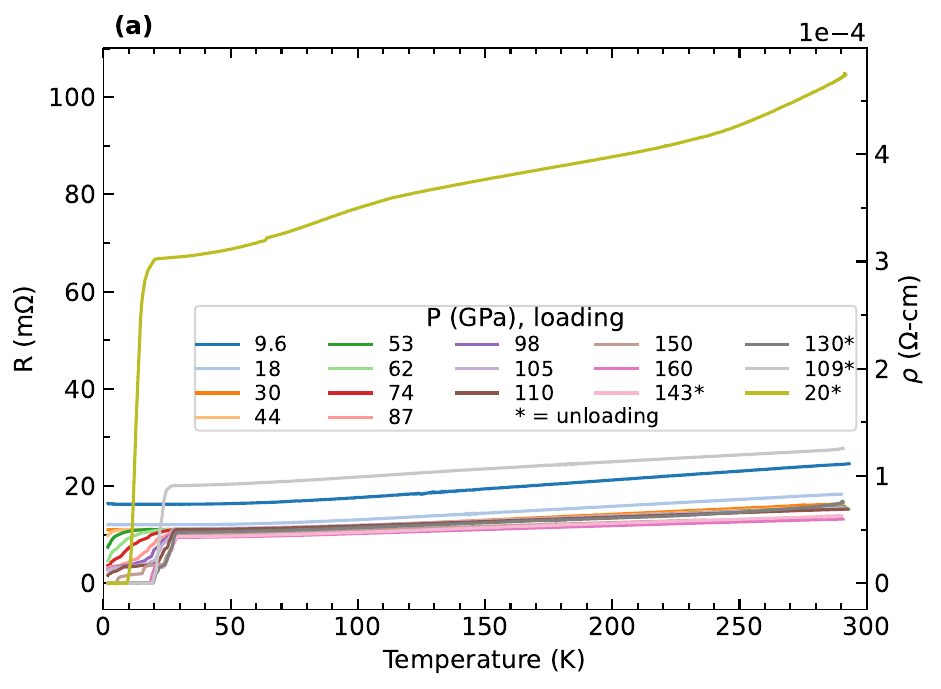}
    \end{subfigure}
    \begin{subfigure}[b]{0.48\textwidth}
        \includegraphics[width=\columnwidth]{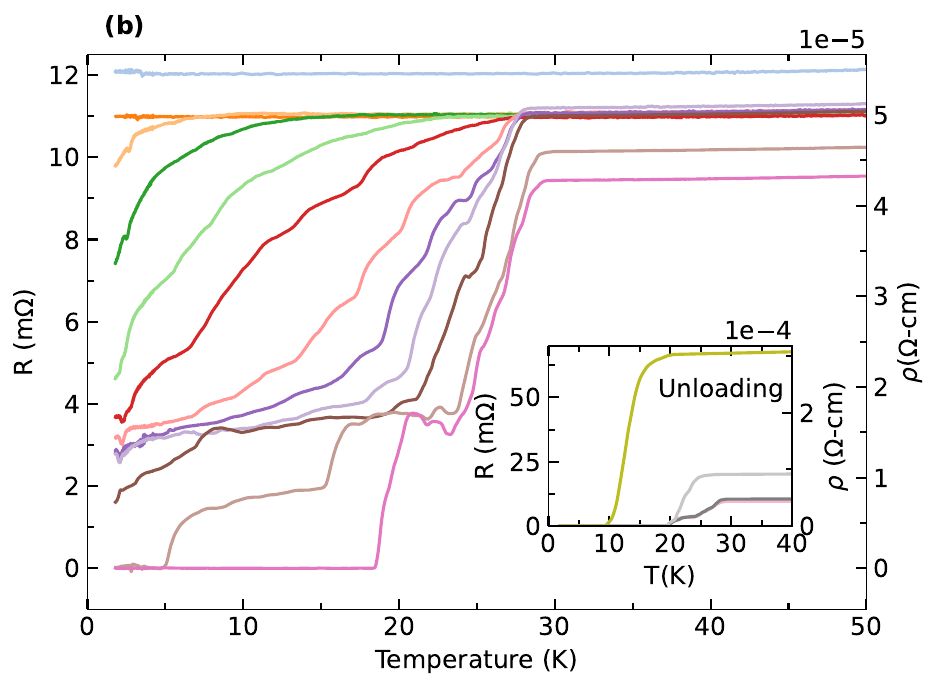}
    \end{subfigure}
    \caption{ (a) Resistance and resistivity dependence on temperature at different pressures. Pressures correspond to the conditions at 10 K.  (b) Zoomed-in plot of the same highlighting the temperature range near \Tc\.  Unloading data are shown in the inset. A drop in the resistance suggesting superconductivity first appears at \SI{44}{GPa} during loading. Resistivity (right y-axis) is a rough estimate from resistance that does not take into account changes in the sample thickness with pressure (see Sec.~\ref{sec:methods}).}
    \label{fig:RvsT}
\end{figure}
\section{Results}
Figure~\ref{fig:RvsT} shows the results of the electrical resistance vs.\ temperature measurements at several pressures.
The sample does not show any sign of superconductivity below \SI{44}{GPa}.
A small drop in resistance is observed at \SI{44}{GPa}, which resembles a partial superconducting transition.
The drop gradually becomes larger with pressure, sometimes occurring in multiple steps, suggesting sample inhomogeneity.
At \SI{150}{GPa} and above, we see a complete superconducting transition with resistance dropping effectively to zero.
Complete transitions are also seen when unloading pressure all the way down to \SI{20}{GPa}.

\begin{figure}[ht]
    \centering
    \includegraphics[width=0.95\columnwidth]{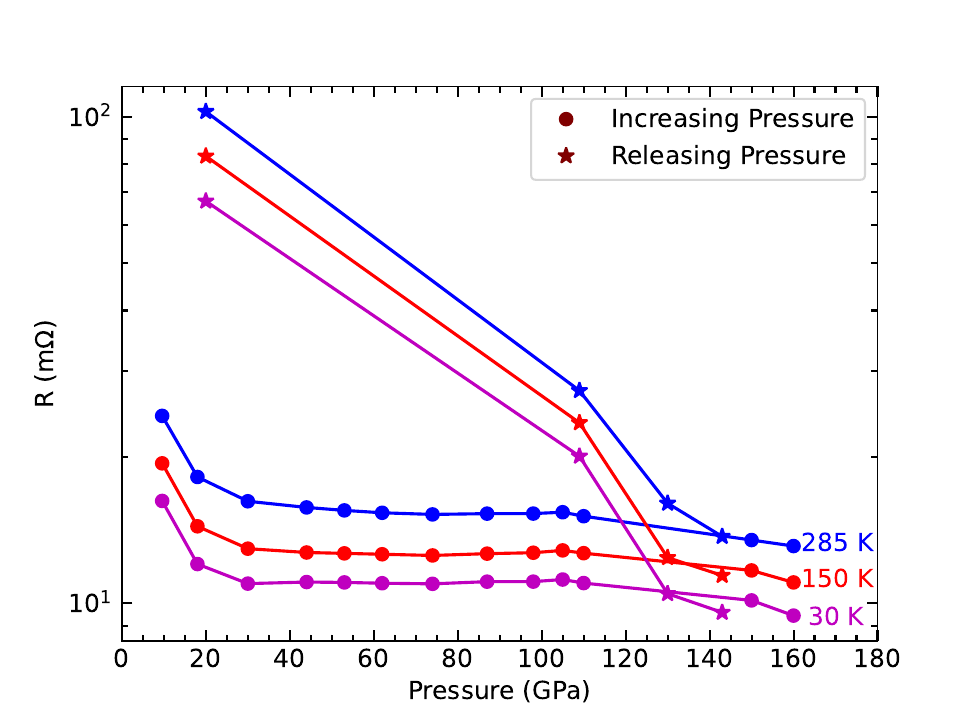}
    \caption{Resistance versus pressure for 285, 150, and \SI{30}{K}.  The data do not exhibit any abrupt features that correspond to the onset of superconductivity and the high pressure \Psmmm\ phase near \SI{45}{GPa}.}
    \label{fig:RvsP}
\end{figure}

The change in resistance with pressure at 285 K, 150 K and 30 K is presented in Fig.~\ref{fig:RvsP}.
The data points are taken from resistance vs temperature measurements at different pressure points.
No sudden change is observed in the resistance vs pressure.
When unloading, the resistance of the sample increases significantly.
This is likely due to the sample being plastically deformed during compression and staying flattened during unloading.

\begin{figure}[ht]
    \centering
    \includegraphics[width=0.95\columnwidth]{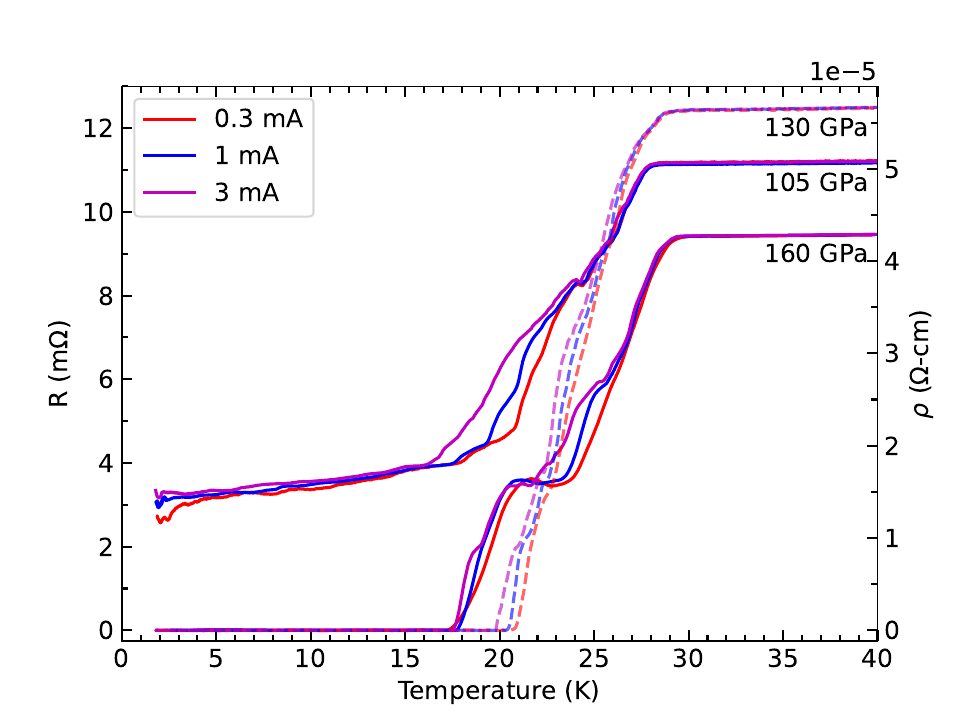}
    \caption{Current dependence measurements performed at three different pressures with excitation currents of 0.3, 1, and 3 mA. The dashed lines represent unloading data (130 GPa). During loading, multiple steps are observed, suggesting significant inhomogeneity.}
    \label{fig:CDM}
\end{figure}

In Fig.~\ref{fig:CDM}, we see the superconducting transition at three different pressure values, 105 and \SI{160}{GPa} (loading), and \SI{130}{GPa} (unloading), recorded with three different excitation currents (0.3, 1, and \SI{3}{mA}).
During loading, the transitions exhibit multiple steps, each of which responds differently to changes in the excitation current.
This supports the conclusion that the sample is inhomogeneous with local variations in the extent to which the superconducting phase percolates ~\cite{multisteptransition,MENEGOTTOCOSTA2013202}.
During unloading, the transition sharpens, and the step-like features become less prominent, responding uniformly to the applied current.
One way to interpret this is that the local defects have spread more uniformly throughout the sample.

\begin{figure*}[t] 
    \centering
    \begin{subfigure}[b]{0.48\textwidth}
        \includegraphics[width=\textwidth]{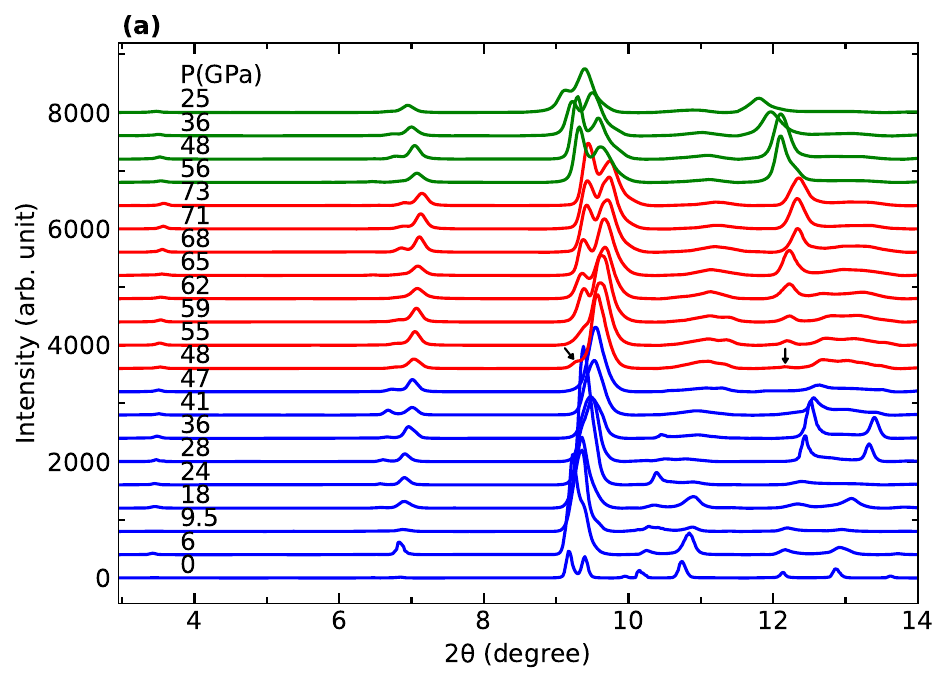}
        \label{fig:XRD_Run1}
    \end{subfigure}
    \hfill
    \begin{subfigure}[b]{0.48\textwidth}
        \includegraphics[width=\textwidth]{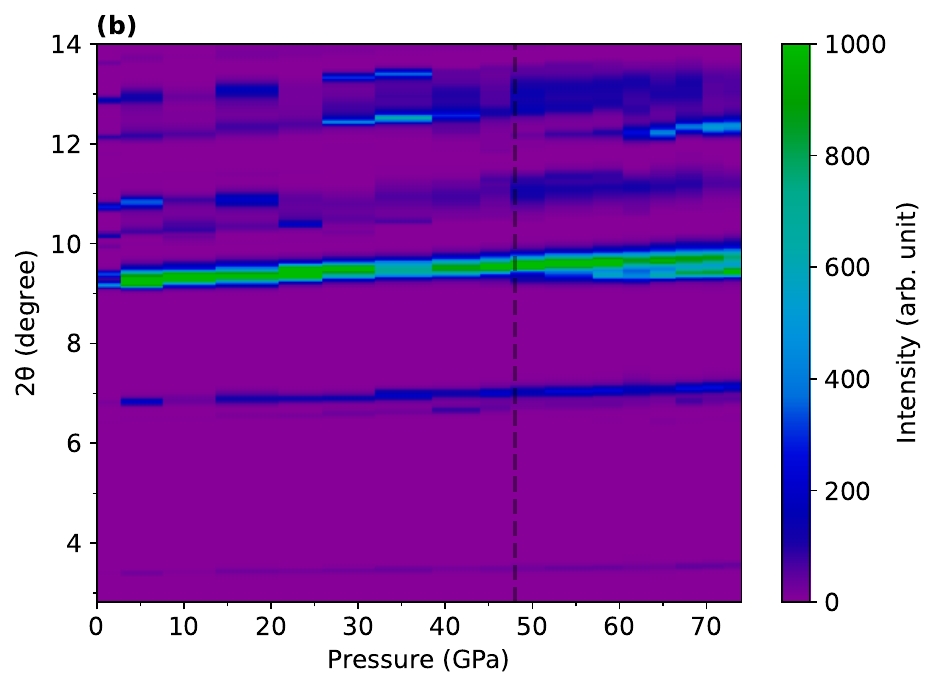}
        \label{fig:Contour_Run1}
    \end{subfigure}
    \vskip\baselineskip
    \begin{subfigure}[b]{0.48\textwidth}
        \includegraphics[width=\textwidth]{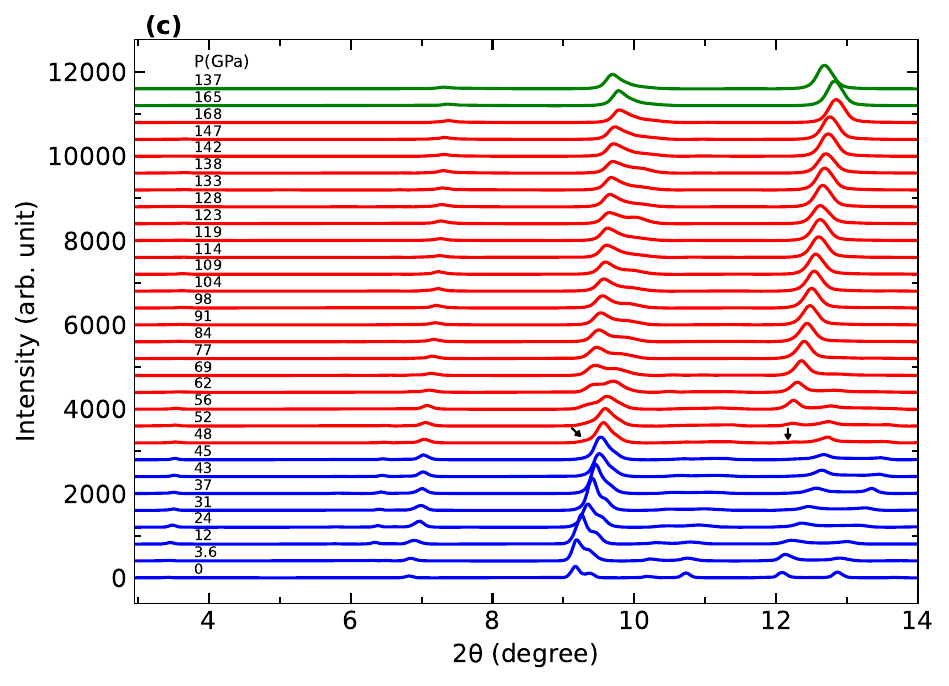}
        \label{fig:Contour_Run2a}
    \end{subfigure}
    \hfill
    \begin{subfigure}[b]{0.48\textwidth}
        \includegraphics[width=\textwidth]{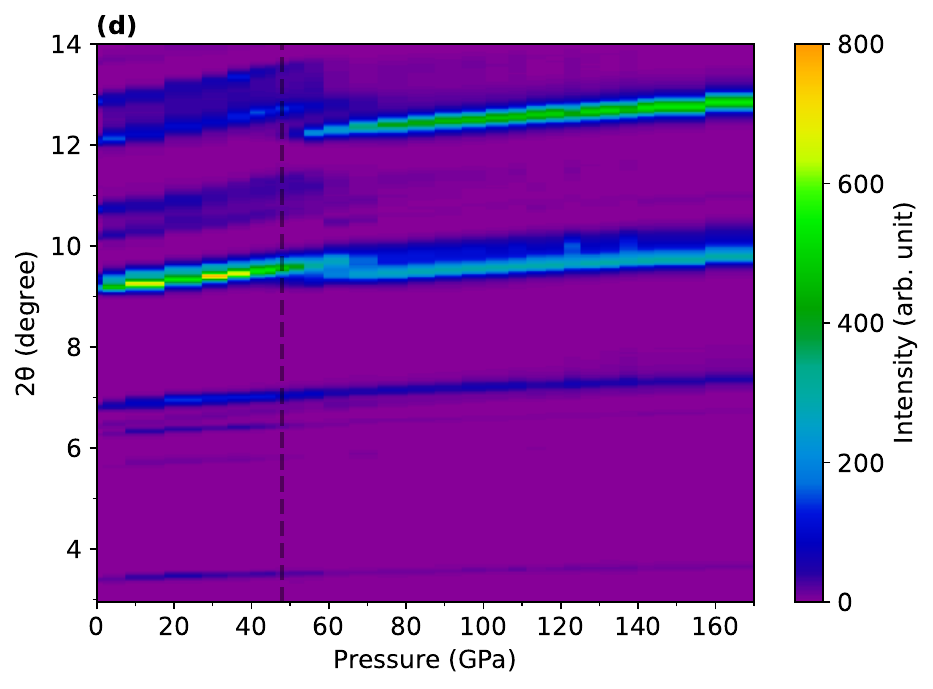}
        \label{fig:Contour_Run2b}
    \end{subfigure}
    \caption{ 
        High pressure XRD measurements performed on \ReMoB.
        (a) 1D XRD patterns from Run 1. 
        The peaks corresponding to the high pressure structure persists down to \SI{25}{GPa} when unloading, (b) corresponding contour plot of XRD peaks with pressure from Run 1,  (c) 1D XRD patterns from Run 2, and (d) corresponding contour plot for Run 2. 
        The 1D XRD curves have been offset vertically for clarity.
        The XRD plots show a structural transition starting around \SI{48}{GPa} (new peaks denoted by black arrows in the 1D XRD plots). 
        The peaks corresponding to the low-pressure structure continue to exist at higher pressure (albeit with lower intensity) indicating a mixed-phase sample.
        In the 1D XRD patterns, blue curves correspond to datasets where the low-pressure phase is dominant, while red curves correspond to  mixed-phase samples. The green curves in XRD were measured during unloading and also show a mixture of phases.
    }
    \label{fig:XRD}
\end{figure*}

\begin{figure}[ht]
\centering
\includegraphics[width=0.48\textwidth]{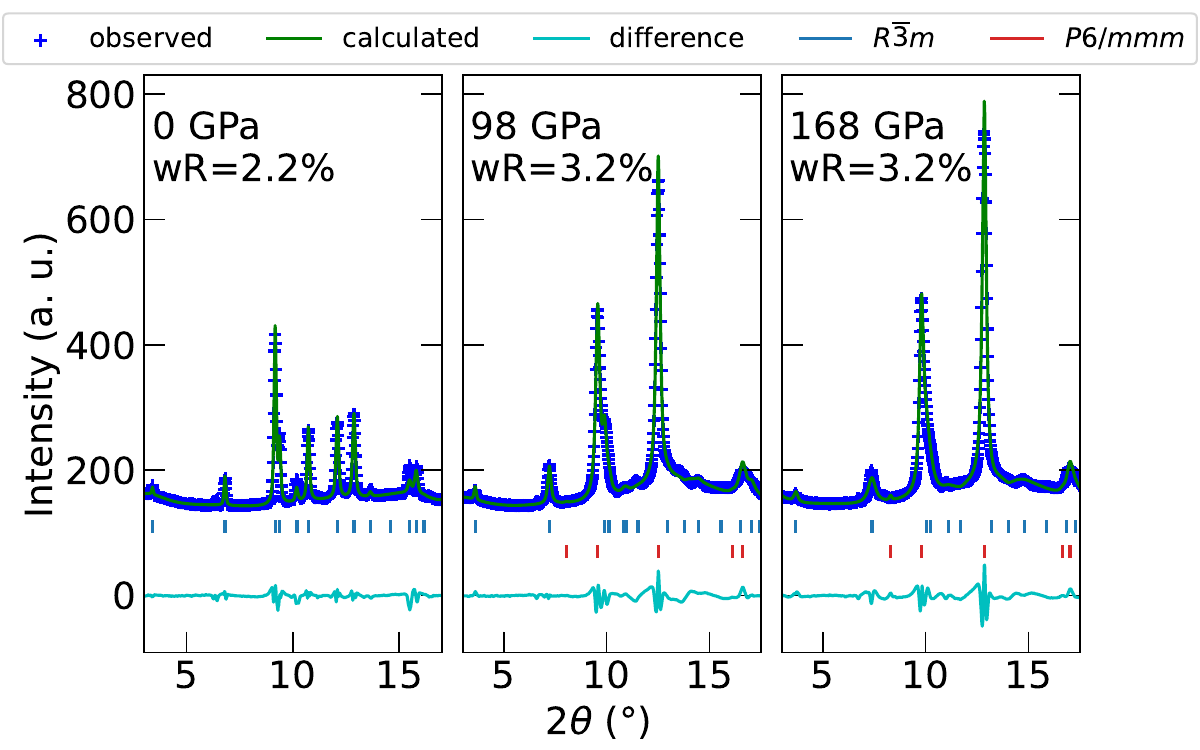}  
\caption{Typical Rietveld refinements on XRD data at three pressures.  Only trace amounts of the \Rtbm\ phase remain at the highest pressure of \SI{168}{GPa}.}
\label{fig:refinement}
\end{figure}

\begin{figure}[ht]
 \centering
\includegraphics[width=0.48\textwidth]{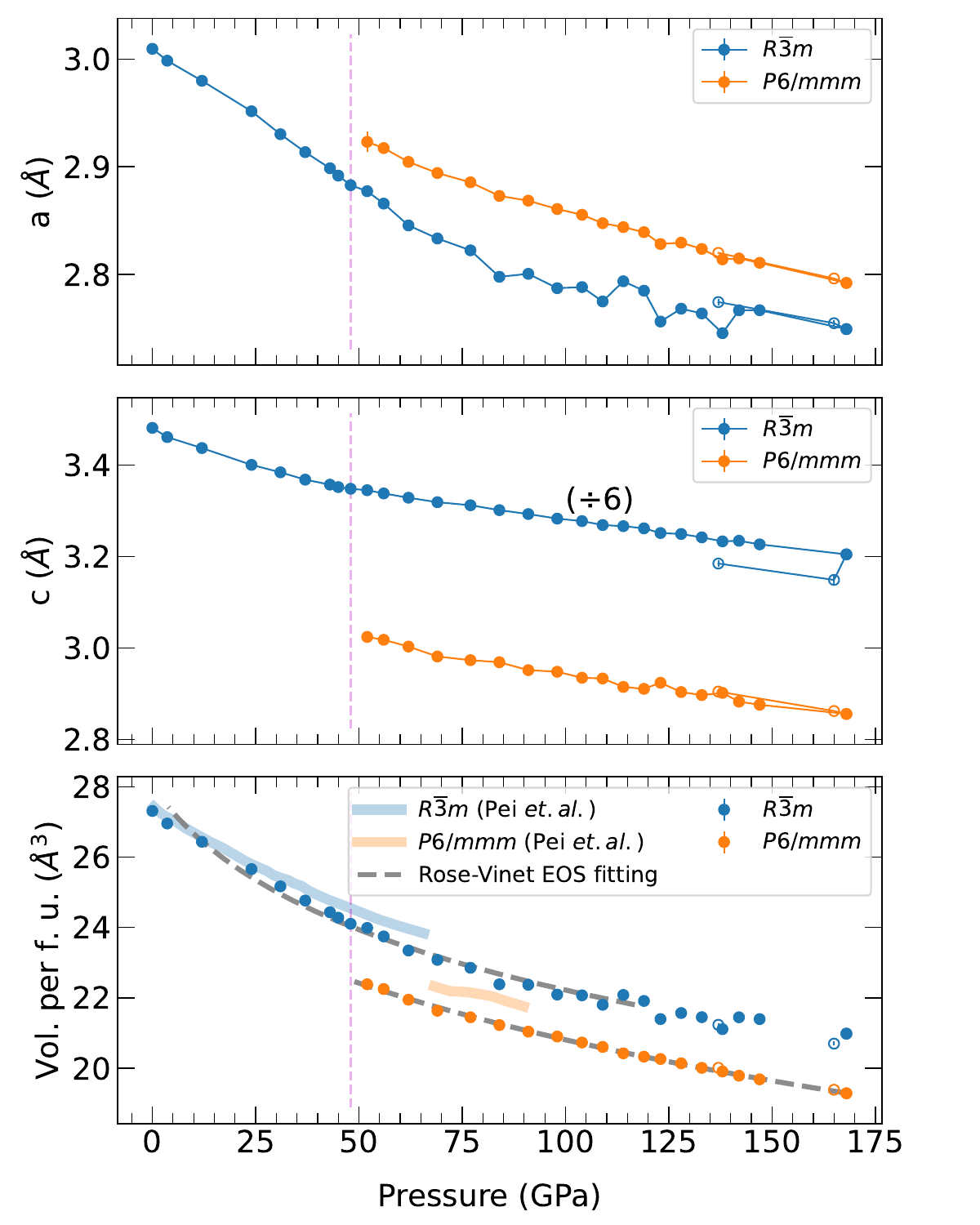} 
    \caption{Pressure dependence of lattice parameters and volume per formula unit. Hollow markers corresponds to unloading. Vertical dashed lines indicate the onset of the structural transition.}
    \label{fig:pressure_depedence}
\end{figure}

X-ray diffraction measurements show the emergence of new peaks beginning at 48 GPa, corresponding to the advent of a structural phase transition from \Rtbm~to \Psmmm\ (see Fig.~\ref{fig:XRD}).
After the commencement of the structural transition at 48 GPa, \ReMoB\ exists in a mixture of the \Rtbm\ and \Psmmm\ structures up to the maximum applied pressure (168 GPa).
The proportion of \Psmmm\ phase increases as the pressure is increased (evident from Fig.~\ref{fig:XRD}d). During unloading, the high pressure phase (\Psmmm) persists down to the lowest pressure (\SI{25}{GPa}) measured during unloading (see Fig.~\ref{fig:XRD}a). 
Rietveld refinements applied to XRD data are presented in Fig.~\ref{fig:refinement} along with the pressure dependence of lattice parameters and volume per formula unit (f.u.) for both phases. 
Similar to \MoB~\cite{MoB2_superconductivity}, the theoretically allowed (001) peak of the \Psmmm\ phase doesn't appear and the volume per f.u.\ drops around 7\% when the 
\Psmmm\ phase transforms into \Psmmm\ phase. 
As shown in Fig.~\ref{fig:pressure_depedence}, Rose-Vinet Equation of State (EOS) fits are performed on the data for the two phases.
The fitting on the \Psmmm\ phase covers pressures only up to 109 GPa since the intensity of the corresponding peaks becomes too small for reliable fitting at higher pressures. The EOS fittings give B\textsubscript{0} = 274(30) GPa, V\textsubscript{0} = 164.73(99)\r{A}$^3$, and B$_0^\prime$ = 4.75(75) for the \Rtbm\ structure and B\textsubscript{0} = 348(52) GPa, V\textsubscript{0} = 25.03(27)\r{A}$^3$ and B$_0^\prime$ = 4.90(69) for the \Psmmm\ structure.

The superconducting phase diagram of \ReMoB\ is shown in Fig.~\ref{fig:PD}. 
The onset \Tc\ increases with pressure above the critical pressure for the structural phase transition and then reaches a maximum value, saturating at $\sim \SI{30}{K}$ near \SI{150}{GPa}.
The \Tc\ values corresponding to loading and unloading agree at relatively high pressure ($>$ 100 GPa). 
However, on unloading, we observe metastability of the superconducting phase to the lowest measured pressure of \SI{20}{GPa}. 
The onset \Tc\ of \SI{20}{K} at \SI{20}{GPa} during unloading contrasts with the lack of a superconducting transition at the same pressure during loading.
The metastability of the superconductivity phase can be explained by the XRD measurements, which show that the high-pressure phase remains metastable down to at least \SI{25}{GPa}.

\section{Discussion}
\begin{figure}[ht]
    \centering
    \includegraphics[width=0.48\textwidth]{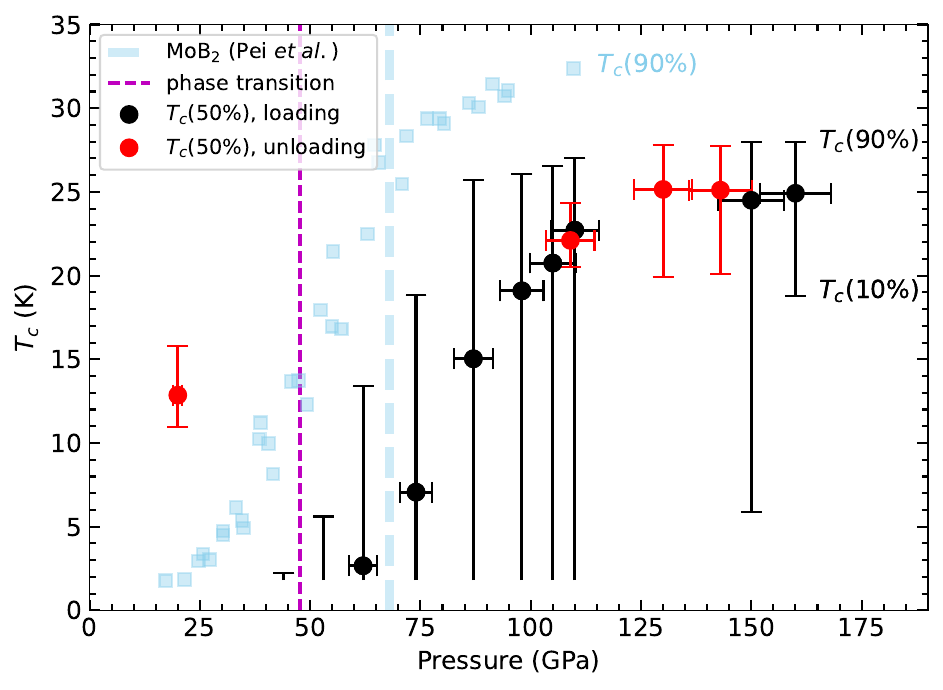}
    \caption{Superconducting phase diagram for \ReMoB\ up to 160 GPa, including both loading and unloading measurements.
    The sample is superconducting at \SI{20}{GPa} when unloading, which shows metastability of the superconducting state.
    The purple and blue dashed lines show the pressure at which structural transition starts in \ReMoB\ and \MoB, respectively.
    Considering both phase diagrams, the kink in \Tc\ vs pressure and the onset of the structural transition appear to be unrelated.}
    \label{fig:PD}
\end{figure}

From the phase diagram of pure \MoB\ alone (see Fig.~\ref{fig:PD}), it may appear that the increase in \Tc\ is characteristic of \Rtbm\ structure and the observed saturation in \Tc~vs pressure is due to structural transition to \Psmmm\ structure.
However, the phase diagram of \ReMoB\ clearly shows that this is not the case.
In contrast to \MoB\, the saturation in \Tc\ is occurs at a higher pressure, while the pressure at which the structure transition starts, is lowered. 
In the case of \ch{WB2}, a similar \Tc\ vs pressure behavior is observed, but no bulk structural transition at all is observed~\cite{Lim_WB2_2021}.
The phase diagram of \ReMoB\  is very similar to both \MoB\ and \ch{WB2} implying that the underlying mechanism for this sharp increase in \Tc\ at lower pressure followed by a flattening of \Tc, is the same in all three cases.
Furthermore, the assertion that the increase in \Tc\ in \MoB\  is characteristic of \Rtbm\ phase is not supported by density functional theory (DFT) calculations~\cite{pei-etal-2023-supplement, quan_private_comm}, which predicted \Rtbm\ structure of \MoB\ to exhibit a very low \Tc\ value ($\sim \SI{5}{K}$) at ambient pressure that further decreases with increase in pressure.
Similar computational results for \ch{WB2} show a vanishing \Tc\ for the low pressure \Psmmc\ structure and a \Tc\ that only decreases with pressure in the \Psmmm\ structure~\cite{Lim_WB2_2021}.
In other words, computational results do not reproduce the portion of the phase diagram where \Tc\ is gradually increasing with pressure for any of the relevant individual crystal structures in these materials.

The initial gradual increase in \Tc~in these materials therefore may be related to a percolation effect~\cite{Stauffer_Aharony_1992, LOBB19781273}.
Given that the material consists of both a metallic phase (\Rtbm) and a superconducting phase (\Psmmm) during the structural transition, the sample should act as a metal-superconductor composite, with resistance and \Tc\ behaviors stemming from percolation and proximity effects.~\cite{osti_6602588, Krivoruchko2019, Sternfeld2005}.
At low proportion of \Psmmm\ structure when this structure has only formed in small disconnected regions of the sample, the grain size of the superconducting phase will be comparable to the superconducting coherence length.
This leads to a suppression in the \Tc\ of the superconducting phase due to inverse proximity effects from the metallic phase~\cite{Deutscher_Gennes_1969,SELEZNYOV2024171645}.


As the pressure increases and the proportion of \Psmmm\ phase increases, larger grains of superconducting phase will form leading to a sharper and larger drop in resistance.
Once the proportion of the \Psmmm\ phase exceeds a critical value such that a continuous superconducting path forms through the sample, a complete superconducting transition with resistance dropping all the way to zero is observed.
Furthermore, when the grain size of the superconducting phase is much larger than the superconducting coherence length, the \Tc\ value will be similar to the bulk superconductor.
When superconductivity first appears at lower pressures—below \SI{48}{GPa} for \ReMoB\ and below \SI{65}{GPa} for \MoB—the grain size of the high-pressure phase is probably too small for XRD detection.
At higher pressures, the bulk behavior takes over and \Tc\ decreases with pressure.
This percolation scenario is consistent with the behavior of the superconducting phase diagram for \ch{WB2}, \MoB, and \ReMoB.

\section{Conclusions}

We studied the effects of Re substitution in \ch{MoB2}. 
Substituting \ch{MoB2} with 10\% Re decreases the pressure corresponding to the onset of a structural phase transition from $\sim$\SI{65}{GPa} down to \SI{48}{GPa}.
However, the pressure at which superconductivity first appears increases from $\sim$ \SI{21}{GPa} in pure \MoB\ to \SI{44}{GPa} in \ReMoB.
The onset \Tc\ for \ReMoB\ gradually increases above \SI{44}{GPa} and reaches a maximum of \SI{30}{K}, compared to $\sim$\SI{32}{K} for \ch{MoB2}. 
Over 90\% of the maximum \Tc\ value is already achieved around 100 GPa.
Our results indicate that the \Psmmm\ structural phase is responsible for the high-temperature superconducting phase in \ReMoB.

The advent of a structural phase transition from \Rtbm\ to \Psmmm\ was suggested as the reason for the sharp change observed in the slope of \Tc\ vs.\ pressure around \SI{65}{GPa} in case of \MoB~\cite{MoB2_superconductivity}.
The present results on \ReMoB\ clearly show that this is not the case since the saturation in \Tc\ occurs around \SI{80}{GPa} while the high-pressure structure is first detected at \SI{48}{GPa} using XRD.
An explanation for the observed change in slope can be obtained by considering percolation effects.
An increase in grain size of the high-pressure superconducting phase with an increase in applied pressure would explain both the observed increase in \Tc\ in the lower pressure range and the saturation of \Tc\ at the highest pressure range.
These grains, if small enough, would not produce detectable XRD peaks.
This effect has also been observed previously in \ch{WB2} where superconductivity appears to occur due to stacking faults resembling the \Psmmm\ structure within a bulk metallic structure (\Psmmc)~\cite{Lim_WB2_2021}.
These results substantiate our previous studies of diborides, supporting the claim that \Psmmm\ phase is crucial to achieving relatively high \Tc\ in diboride superconductors.

\section{Data Availability}
All data and analysis code associated with this study is publicly available~\cite{hamlin_2024_13359794}.
\vspace{1em}

\section{Acknowledgments}
Work at the University of Florida was performed under the auspices of U.S. Department of Energy Basic Energy Sciences under Contract No.\ DE-SC-0020385 and the U.S. National Science Foundation, Division of Materials Research under Contract No.\ NSF-DMR-2118718.
A.C.H.\ and R.G.H.\ acknowledge additional support from the National Science Foundation under award PHY-1549132 (Center for Bright Beams).
R.S.K.\ and R.J.H.\ acknowledge support from the U.S.\ National Science Foundation (DMR-2119308 and DMR-2104881).
X-ray diffraction measurements were performed at HPCAT (Sector 16), Advanced Photon Source (APS), Argonne National Laboratory. HPCAT operations are supported by the DOE-National Nuclear Security Administration (NNSA) Office of Experimental Sciences.
The beamtime was made possible by the Chicago/DOE Alliance Center (CDAC), which is supported by DOE-NNSA (DE-NA0003975).
Use of the gas loading system was supported by COMPRES under NSF Cooperative Agreement EAR-1606856 and by GSECARS through NSF grant EAR-1634415 and DOE grant DE-FG02-94ER14466.
The Advanced Photon Source is a DOE Office of Science User Facility operated for the DOE Office of Science by Argonne National Laboratory under Contract No. DE-AC02-06CH11357.
High pressure equipment development at the University of Florida was supported by National Science Foundation CAREER award DMR-1453752.


\begin{thebibliography}{38}%
\makeatletter
\providecommand \@ifxundefined [1]{%
 \@ifx{#1\undefined}
}%
\providecommand \@ifnum [1]{%
 \ifnum #1\expandafter \@firstoftwo
 \else \expandafter \@secondoftwo
 \fi
}%
\providecommand \@ifx [1]{%
 \ifx #1\expandafter \@firstoftwo
 \else \expandafter \@secondoftwo
 \fi
}%
\providecommand \natexlab [1]{#1}%
\providecommand \enquote  [1]{``#1''}%
\providecommand \bibnamefont  [1]{#1}%
\providecommand \bibfnamefont [1]{#1}%
\providecommand \citenamefont [1]{#1}%
\providecommand \href@noop [0]{\@secondoftwo}%
\providecommand \href [0]{\begingroup \@sanitize@url \@href}%
\providecommand \@href[1]{\@@startlink{#1}\@@href}%
\providecommand \@@href[1]{\endgroup#1\@@endlink}%
\providecommand \@sanitize@url [0]{\catcode `\\12\catcode `\$12\catcode
  `\&12\catcode `\#12\catcode `\^12\catcode `\_12\catcode `\%12\relax}%
\providecommand \@@startlink[1]{}%
\providecommand \@@endlink[0]{}%
\providecommand \url  [0]{\begingroup\@sanitize@url \@url }%
\providecommand \@url [1]{\endgroup\@href {#1}{\urlprefix }}%
\providecommand \urlprefix  [0]{URL }%
\providecommand \Eprint [0]{\href }%
\providecommand \doibase [0]{https://doi.org/}%
\providecommand \selectlanguage [0]{\@gobble}%
\providecommand \bibinfo  [0]{\@secondoftwo}%
\providecommand \bibfield  [0]{\@secondoftwo}%
\providecommand \translation [1]{[#1]}%
\providecommand \BibitemOpen [0]{}%
\providecommand \bibitemStop [0]{}%
\providecommand \bibitemNoStop [0]{.\EOS\space}%
\providecommand \EOS [0]{\spacefactor3000\relax}%
\providecommand \BibitemShut  [1]{\csname bibitem#1\endcsname}%
\let\auto@bib@innerbib\@empty
\bibitem [{\citenamefont {Drozdov}\ \emph {et~al.}(2015)\citenamefont
  {Drozdov}, \citenamefont {Eremets}, \citenamefont {Troyan}, \citenamefont
  {Ksenofontov},\ and\ \citenamefont {Shylin}}]{Drozdov2015}%
  \BibitemOpen
  \bibfield  {author} {\bibinfo {author} {\bibfnamefont {A.~P.}\ \bibnamefont
  {Drozdov}}, \bibinfo {author} {\bibfnamefont {M.~I.}\ \bibnamefont
  {Eremets}}, \bibinfo {author} {\bibfnamefont {I.~A.}\ \bibnamefont {Troyan}},
  \bibinfo {author} {\bibfnamefont {V.}~\bibnamefont {Ksenofontov}},\ and\
  \bibinfo {author} {\bibfnamefont {S.~I.}\ \bibnamefont {Shylin}},\ }\href
  {https://doi.org/10.1038/nature14964} {\bibfield  {journal} {\bibinfo
  {journal} {Nature}\ }\textbf {\bibinfo {volume} {525}},\ \bibinfo {pages}
  {73} (\bibinfo {year} {2015})}\BibitemShut {NoStop}%
\bibitem [{\citenamefont {Papaconstantopoulos}\ \emph
  {et~al.}(2020)\citenamefont {Papaconstantopoulos}, \citenamefont {Mehl},\
  and\ \citenamefont {Chang}}]{LaH10}%
  \BibitemOpen
  \bibfield  {author} {\bibinfo {author} {\bibfnamefont {D.~A.}\ \bibnamefont
  {Papaconstantopoulos}}, \bibinfo {author} {\bibfnamefont {M.~J.}\
  \bibnamefont {Mehl}},\ and\ \bibinfo {author} {\bibfnamefont {P.-H.}\
  \bibnamefont {Chang}},\ }\href {https://doi.org/10.1103/PhysRevB.101.060506}
  {\bibfield  {journal} {\bibinfo  {journal} {Phys. Rev. B}\ }\textbf {\bibinfo
  {volume} {101}},\ \bibinfo {pages} {060506} (\bibinfo {year}
  {2020})}\BibitemShut {NoStop}%
\bibitem [{\citenamefont {Mozaffari}\ \emph {et~al.}(2019)\citenamefont
  {Mozaffari}, \citenamefont {Sun}, \citenamefont {Minkov}, \citenamefont
  {Drozdov}, \citenamefont {Knyazev}, \citenamefont {Betts}, \citenamefont
  {Einaga}, \citenamefont {Shimizu}, \citenamefont {Eremets}, \citenamefont
  {Balicas},\ and\ \citenamefont {Balakirev}}]{Mozaffari2019}%
  \BibitemOpen
  \bibfield  {author} {\bibinfo {author} {\bibfnamefont {S.}~\bibnamefont
  {Mozaffari}}, \bibinfo {author} {\bibfnamefont {D.}~\bibnamefont {Sun}},
  \bibinfo {author} {\bibfnamefont {V.~S.}\ \bibnamefont {Minkov}}, \bibinfo
  {author} {\bibfnamefont {A.~P.}\ \bibnamefont {Drozdov}}, \bibinfo {author}
  {\bibfnamefont {D.}~\bibnamefont {Knyazev}}, \bibinfo {author} {\bibfnamefont
  {J.~B.}\ \bibnamefont {Betts}}, \bibinfo {author} {\bibfnamefont
  {M.}~\bibnamefont {Einaga}}, \bibinfo {author} {\bibfnamefont
  {K.}~\bibnamefont {Shimizu}}, \bibinfo {author} {\bibfnamefont {M.~I.}\
  \bibnamefont {Eremets}}, \bibinfo {author} {\bibfnamefont {L.}~\bibnamefont
  {Balicas}},\ and\ \bibinfo {author} {\bibfnamefont {F.~F.}\ \bibnamefont
  {Balakirev}},\ }\href {https://doi.org/10.1038/s41467-019-10552-y} {\bibfield
   {journal} {\bibinfo  {journal} {Nature Communications}\ }\textbf {\bibinfo
  {volume} {10}},\ \bibinfo {pages} {2522} (\bibinfo {year}
  {2019})}\BibitemShut {NoStop}%
\bibitem [{\citenamefont {An}\ and\ \citenamefont
  {Pickett}(2001)}]{AnPickett2001}%
  \BibitemOpen
  \bibfield  {author} {\bibinfo {author} {\bibfnamefont {J.~M.}\ \bibnamefont
  {An}}\ and\ \bibinfo {author} {\bibfnamefont {W.~E.}\ \bibnamefont
  {Pickett}},\ }\href {https://doi.org/10.1103/PhysRevLett.86.4366} {\bibfield
  {journal} {\bibinfo  {journal} {Phys. Rev. Lett.}\ }\textbf {\bibinfo
  {volume} {86}},\ \bibinfo {pages} {4366} (\bibinfo {year}
  {2001})}\BibitemShut {NoStop}%
\bibitem [{\citenamefont {Mazin}\ and\ \citenamefont
  {Antropov}(2003)}]{Mazin2003}%
  \BibitemOpen
  \bibfield  {author} {\bibinfo {author} {\bibfnamefont {I.}~\bibnamefont
  {Mazin}}\ and\ \bibinfo {author} {\bibfnamefont {V.}~\bibnamefont
  {Antropov}},\ }\href
  {https://doi.org/https://doi.org/10.1016/S0921-4534(02)02299-2} {\bibfield
  {journal} {\bibinfo  {journal} {Physica C: Superconductivity}\ }\textbf
  {\bibinfo {volume} {385}},\ \bibinfo {pages} {49} (\bibinfo {year}
  {2003})}\BibitemShut {NoStop}%
\bibitem [{\citenamefont {Buzea}\ and\ \citenamefont
  {Yamashita}(2001)}]{buzea_review_2001}%
  \BibitemOpen
  \bibfield  {author} {\bibinfo {author} {\bibfnamefont {C.}~\bibnamefont
  {Buzea}}\ and\ \bibinfo {author} {\bibfnamefont {T.}~\bibnamefont
  {Yamashita}},\ }\href {https://doi.org/10.1088/0953-2048/14/11/201}
  {\bibfield  {journal} {\bibinfo  {journal} {Superconductor Science and
  Technology}\ }\textbf {\bibinfo {volume} {14}},\ \bibinfo {pages} {R115}
  (\bibinfo {year} {2001})}\BibitemShut {NoStop}%
\bibitem [{\citenamefont {Bud’ko}\ and\ \citenamefont
  {Canfield}(2015{\natexlab{a}})}]{budko_superconductivity_2015}%
  \BibitemOpen
  \bibfield  {author} {\bibinfo {author} {\bibfnamefont {S.~L.}\ \bibnamefont
  {Bud’ko}}\ and\ \bibinfo {author} {\bibfnamefont {P.~C.}\ \bibnamefont
  {Canfield}},\ }\href {https://doi.org/10.1016/j.physc.2015.02.024} {\bibfield
   {journal} {\bibinfo  {journal} {Physica C: Superconductivity and its
  Applications}\ }\bibinfo {series} {Superconducting {Materials}:
  {Conventional}, {Unconventional} and {Undetermined}},\ \textbf {\bibinfo
  {volume} {514}},\ \bibinfo {pages} {142} (\bibinfo {year}
  {2015}{\natexlab{a}})}\BibitemShut {NoStop}%
\bibitem [{\citenamefont {Carvalho~de
  Castro~Sene}(2024)}]{CarvalhodeCastroSene2024}%
  \BibitemOpen
  \bibfield  {author} {\bibinfo {author} {\bibfnamefont {F.}~\bibnamefont
  {Carvalho~de Castro~Sene}},\ }\href
  {https://doi.org/10.1016/j.supcon.2023.100083} {\bibfield  {journal}
  {\bibinfo  {journal} {Superconductivity}\ }\textbf {\bibinfo {volume} {9}},\
  \bibinfo {pages} {100083} (\bibinfo {year} {2024})}\BibitemShut {NoStop}%
\bibitem [{\citenamefont {Pei}\ \emph {et~al.}()\citenamefont {Pei},
  \citenamefont {Zhang}, \citenamefont {Wang}, \citenamefont {Zhao},
  \citenamefont {Gao}, \citenamefont {Gong}, \citenamefont {Tian},
  \citenamefont {Luo}, \citenamefont {Li}, \citenamefont {Yang}, \citenamefont
  {Lu}, \citenamefont {Lei}, \citenamefont {Liu},\ and\ \citenamefont
  {Qi}}]{MoB2_superconductivity}%
  \BibitemOpen
  \bibfield  {author} {\bibinfo {author} {\bibfnamefont {C.}~\bibnamefont
  {Pei}}, \bibinfo {author} {\bibfnamefont {J.}~\bibnamefont {Zhang}}, \bibinfo
  {author} {\bibfnamefont {Q.}~\bibnamefont {Wang}}, \bibinfo {author}
  {\bibfnamefont {Y.}~\bibnamefont {Zhao}}, \bibinfo {author} {\bibfnamefont
  {L.}~\bibnamefont {Gao}}, \bibinfo {author} {\bibfnamefont {C.}~\bibnamefont
  {Gong}}, \bibinfo {author} {\bibfnamefont {S.}~\bibnamefont {Tian}}, \bibinfo
  {author} {\bibfnamefont {R.}~\bibnamefont {Luo}}, \bibinfo {author}
  {\bibfnamefont {M.}~\bibnamefont {Li}}, \bibinfo {author} {\bibfnamefont
  {W.}~\bibnamefont {Yang}}, \bibinfo {author} {\bibfnamefont {Z.-Y.}\
  \bibnamefont {Lu}}, \bibinfo {author} {\bibfnamefont {H.}~\bibnamefont
  {Lei}}, \bibinfo {author} {\bibfnamefont {K.}~\bibnamefont {Liu}},\ and\
  \bibinfo {author} {\bibfnamefont {Y.}~\bibnamefont {Qi}},\ }\href
  {https://doi.org/10.1093/nsr/nwad034} {\bibinfo  {journal} {Natl. Sci. Rev.,
  nwad034 (2023)}\ }\BibitemShut {NoStop}%
\bibitem [{\citenamefont {Quan}\ \emph {et~al.}(2021)\citenamefont {Quan},
  \citenamefont {Lee},\ and\ \citenamefont {Pickett}}]{Quan2021}%
  \BibitemOpen
\bibfield  {journal} {  }\bibfield  {author} {\bibinfo {author} {\bibfnamefont
  {Y.}~\bibnamefont {Quan}}, \bibinfo {author} {\bibfnamefont {K.-W.}\
  \bibnamefont {Lee}},\ and\ \bibinfo {author} {\bibfnamefont {W.~E.}\
  \bibnamefont {Pickett}},\ }\href
  {https://doi.org/10.1103/PhysRevB.104.224504} {\bibfield  {journal} {\bibinfo
   {journal} {Phys. Rev. B}\ }\textbf {\bibinfo {volume} {104}},\ \bibinfo
  {pages} {224504} (\bibinfo {year} {2021})}\BibitemShut {NoStop}%
\bibitem [{\citenamefont {Bud’ko}\ and\ \citenamefont
  {Canfield}(2015{\natexlab{b}})}]{BUDKO2015142}%
  \BibitemOpen
  \bibfield  {author} {\bibinfo {author} {\bibfnamefont {S.~L.}\ \bibnamefont
  {Bud’ko}}\ and\ \bibinfo {author} {\bibfnamefont {P.~C.}\ \bibnamefont
  {Canfield}},\ }\href
  {https://doi.org/https://doi.org/10.1016/j.physc.2015.02.024} {\bibfield
  {journal} {\bibinfo  {journal} {Physica C: Superconductivity and its
  Applications}\ }\textbf {\bibinfo {volume} {514}},\ \bibinfo {pages} {142}
  (\bibinfo {year} {2015}{\natexlab{b}})},\ \bibinfo {note} {superconducting
  Materials: Conventional, Unconventional and Undetermined}\BibitemShut
  {NoStop}%
\bibitem [{\citenamefont {Hinks}\ \emph {et~al.}(2001)\citenamefont {Hinks},
  \citenamefont {Claus},\ and\ \citenamefont {Jorgensen}}]{Hinks2001}%
  \BibitemOpen
  \bibfield  {author} {\bibinfo {author} {\bibfnamefont {D.~G.}\ \bibnamefont
  {Hinks}}, \bibinfo {author} {\bibfnamefont {H.}~\bibnamefont {Claus}},\ and\
  \bibinfo {author} {\bibfnamefont {J.~D.}\ \bibnamefont {Jorgensen}},\ }\href
  {https://doi.org/10.1038/35078037} {\bibfield  {journal} {\bibinfo  {journal}
  {Nature}\ }\textbf {\bibinfo {volume} {411}},\ \bibinfo {pages} {457}
  (\bibinfo {year} {2001})}\BibitemShut {NoStop}%
\bibitem [{\citenamefont {Lim}\ \emph {et~al.}(2022)\citenamefont {Lim},
  \citenamefont {Hire}, \citenamefont {Quan}, \citenamefont {Kim},
  \citenamefont {Xie}, \citenamefont {Sinha}, \citenamefont {Kumar},
  \citenamefont {Popov}, \citenamefont {Park}, \citenamefont {Hemley},
  \citenamefont {Vohra}, \citenamefont {Hamlin}, \citenamefont {Hennig},
  \citenamefont {Hirschfeld},\ and\ \citenamefont {Stewart}}]{Lim_WB2_2021}%
  \BibitemOpen
  \bibfield  {author} {\bibinfo {author} {\bibfnamefont {J.}~\bibnamefont
  {Lim}}, \bibinfo {author} {\bibfnamefont {A.~C.}\ \bibnamefont {Hire}},
  \bibinfo {author} {\bibfnamefont {Y.}~\bibnamefont {Quan}}, \bibinfo {author}
  {\bibfnamefont {J.~S.}\ \bibnamefont {Kim}}, \bibinfo {author} {\bibfnamefont
  {S.~R.}\ \bibnamefont {Xie}}, \bibinfo {author} {\bibfnamefont
  {S.}~\bibnamefont {Sinha}}, \bibinfo {author} {\bibfnamefont {R.~S.}\
  \bibnamefont {Kumar}}, \bibinfo {author} {\bibfnamefont {D.}~\bibnamefont
  {Popov}}, \bibinfo {author} {\bibfnamefont {C.}~\bibnamefont {Park}},
  \bibinfo {author} {\bibfnamefont {R.~J.}\ \bibnamefont {Hemley}}, \bibinfo
  {author} {\bibfnamefont {Y.~K.}\ \bibnamefont {Vohra}}, \bibinfo {author}
  {\bibfnamefont {J.~J.}\ \bibnamefont {Hamlin}}, \bibinfo {author}
  {\bibfnamefont {R.~G.}\ \bibnamefont {Hennig}}, \bibinfo {author}
  {\bibfnamefont {P.~J.}\ \bibnamefont {Hirschfeld}},\ and\ \bibinfo {author}
  {\bibfnamefont {G.~R.}\ \bibnamefont {Stewart}},\ }\href
  {https://doi.org/https://doi.org/10.1038/s41467-022-35191-8} {\bibfield
  {journal} {\bibinfo  {journal} {Nat. Commun.}\ }\textbf {\bibinfo {volume}
  {13}},\ \bibinfo {pages} {7901} (\bibinfo {year} {2022})}\BibitemShut
  {NoStop}%
\bibitem [{\citenamefont {Hire}\ \emph {et~al.}(2022)\citenamefont {Hire},
  \citenamefont {Sinha}, \citenamefont {Lim}, \citenamefont {Kim},
  \citenamefont {Dee}, \citenamefont {Fanfarillo}, \citenamefont {Hamlin},
  \citenamefont {Hennig}, \citenamefont {Hirschfeld},\ and\ \citenamefont
  {Stewart}}]{Hire_NbxMoxB2_2022}%
  \BibitemOpen
  \bibfield  {author} {\bibinfo {author} {\bibfnamefont {A.~C.}\ \bibnamefont
  {Hire}}, \bibinfo {author} {\bibfnamefont {S.}~\bibnamefont {Sinha}},
  \bibinfo {author} {\bibfnamefont {J.}~\bibnamefont {Lim}}, \bibinfo {author}
  {\bibfnamefont {J.~S.}\ \bibnamefont {Kim}}, \bibinfo {author} {\bibfnamefont
  {P.~M.}\ \bibnamefont {Dee}}, \bibinfo {author} {\bibfnamefont
  {L.}~\bibnamefont {Fanfarillo}}, \bibinfo {author} {\bibfnamefont {J.~J.}\
  \bibnamefont {Hamlin}}, \bibinfo {author} {\bibfnamefont {R.~G.}\
  \bibnamefont {Hennig}}, \bibinfo {author} {\bibfnamefont {P.~J.}\
  \bibnamefont {Hirschfeld}},\ and\ \bibinfo {author} {\bibfnamefont {G.~R.}\
  \bibnamefont {Stewart}},\ }\href
  {https://doi.org/10.1103/PhysRevB.106.174515} {\bibfield  {journal} {\bibinfo
   {journal} {Phys. Rev. B}\ }\textbf {\bibinfo {volume} {106}},\ \bibinfo
  {pages} {174515} (\bibinfo {year} {2022})}\BibitemShut {NoStop}%
\bibitem [{\citenamefont {Lim}\ \emph {et~al.}(2023)\citenamefont {Lim},
  \citenamefont {Sinha}, \citenamefont {Hire}, \citenamefont {Kim},
  \citenamefont {Dee}, \citenamefont {Kumar}, \citenamefont {Popov},
  \citenamefont {Hemley}, \citenamefont {Hennig}, \citenamefont {Hirschfeld},
  \citenamefont {Stewart},\ and\ \citenamefont {Hamlin}}]{Lim2023}%
  \BibitemOpen
  \bibfield  {author} {\bibinfo {author} {\bibfnamefont {J.}~\bibnamefont
  {Lim}}, \bibinfo {author} {\bibfnamefont {S.}~\bibnamefont {Sinha}}, \bibinfo
  {author} {\bibfnamefont {A.~C.}\ \bibnamefont {Hire}}, \bibinfo {author}
  {\bibfnamefont {J.~S.}\ \bibnamefont {Kim}}, \bibinfo {author} {\bibfnamefont
  {P.~M.}\ \bibnamefont {Dee}}, \bibinfo {author} {\bibfnamefont {R.~S.}\
  \bibnamefont {Kumar}}, \bibinfo {author} {\bibfnamefont {D.}~\bibnamefont
  {Popov}}, \bibinfo {author} {\bibfnamefont {R.~J.}\ \bibnamefont {Hemley}},
  \bibinfo {author} {\bibfnamefont {R.~G.}\ \bibnamefont {Hennig}}, \bibinfo
  {author} {\bibfnamefont {P.~J.}\ \bibnamefont {Hirschfeld}}, \bibinfo
  {author} {\bibfnamefont {G.~R.}\ \bibnamefont {Stewart}},\ and\ \bibinfo
  {author} {\bibfnamefont {J.~J.}\ \bibnamefont {Hamlin}},\ }\href
  {https://doi.org/10.1103/PhysRevB.108.094501} {\bibfield  {journal} {\bibinfo
   {journal} {Phys. Rev. B}\ }\textbf {\bibinfo {volume} {108}},\ \bibinfo
  {pages} {094501} (\bibinfo {year} {2023})}\BibitemShut {NoStop}%
\bibitem [{\citenamefont {Dee}\ \emph {et~al.}(2024)\citenamefont {Dee},
  \citenamefont {Kim}, \citenamefont {Hire}, \citenamefont {Lim}, \citenamefont
  {Fanfarillo}, \citenamefont {Sinha}, \citenamefont {Hamlin}, \citenamefont
  {Hennig}, \citenamefont {Hirschfeld},\ and\ \citenamefont
  {Stewart}}]{Dee2024}%
  \BibitemOpen
  \bibfield  {author} {\bibinfo {author} {\bibfnamefont {P.~M.}\ \bibnamefont
  {Dee}}, \bibinfo {author} {\bibfnamefont {J.~S.}\ \bibnamefont {Kim}},
  \bibinfo {author} {\bibfnamefont {A.~C.}\ \bibnamefont {Hire}}, \bibinfo
  {author} {\bibfnamefont {J.}~\bibnamefont {Lim}}, \bibinfo {author}
  {\bibfnamefont {L.}~\bibnamefont {Fanfarillo}}, \bibinfo {author}
  {\bibfnamefont {S.}~\bibnamefont {Sinha}}, \bibinfo {author} {\bibfnamefont
  {J.~J.}\ \bibnamefont {Hamlin}}, \bibinfo {author} {\bibfnamefont {R.~G.}\
  \bibnamefont {Hennig}}, \bibinfo {author} {\bibfnamefont {P.~J.}\
  \bibnamefont {Hirschfeld}},\ and\ \bibinfo {author} {\bibfnamefont {G.~R.}\
  \bibnamefont {Stewart}},\ }\href
  {https://doi.org/10.1103/PhysRevB.109.104520} {\bibfield  {journal} {\bibinfo
   {journal} {Phys. Rev. B}\ }\textbf {\bibinfo {volume} {109}},\ \bibinfo
  {pages} {104520} (\bibinfo {year} {2024})}\BibitemShut {NoStop}%
\bibitem [{\citenamefont {Clementi}\ \emph {et~al.}(1967)\citenamefont
  {Clementi}, \citenamefont {Raimondi},\ and\ \citenamefont
  {Reinhardt}}]{clementi1967atomic}%
  \BibitemOpen
  \bibfield  {author} {\bibinfo {author} {\bibfnamefont {E.}~\bibnamefont
  {Clementi}}, \bibinfo {author} {\bibfnamefont {D.~L.}\ \bibnamefont
  {Raimondi}},\ and\ \bibinfo {author} {\bibfnamefont {W.~P.}\ \bibnamefont
  {Reinhardt}},\ }\href {https://doi.org/10.1063/1.1712084} {\bibfield
  {journal} {\bibinfo  {journal} {The Journal of Chemical Physics}\ }\textbf
  {\bibinfo {volume} {47}},\ \bibinfo {pages} {1300} (\bibinfo {year}
  {1967})}\BibitemShut {NoStop}%
\bibitem [{\citenamefont {Wang}\ \emph {et~al.}(2019)\citenamefont {Wang},
  \citenamefont {Yan}, \citenamefont {Liu},\ and\ \citenamefont
  {Zhou}}]{RheniumDiboride}%
  \BibitemOpen
  \bibfield  {author} {\bibinfo {author} {\bibfnamefont {Y.~X.}\ \bibnamefont
  {Wang}}, \bibinfo {author} {\bibfnamefont {Z.~X.}\ \bibnamefont {Yan}},
  \bibinfo {author} {\bibfnamefont {W.}~\bibnamefont {Liu}},\ and\ \bibinfo
  {author} {\bibfnamefont {G.~L.}\ \bibnamefont {Zhou}},\ }\href
  {https://doi.org/10.1063/1.5115245} {\bibfield  {journal} {\bibinfo
  {journal} {Journal of Applied Physics}\ }\textbf {\bibinfo {volume} {126}},\
  \bibinfo {pages} {135901} (\bibinfo {year} {2019})}\BibitemShut {NoStop}%
\bibitem [{\citenamefont {Jain}\ \emph {et~al.}(2013)\citenamefont {Jain},
  \citenamefont {Ong}, \citenamefont {Hautier}, \citenamefont {Chen},
  \citenamefont {Richards}, \citenamefont {Dacek}, \citenamefont {Cholia},
  \citenamefont {Gunter}, \citenamefont {Skinner}, \citenamefont {Ceder},\ and\
  \citenamefont {Persson}}]{Jain2013}%
  \BibitemOpen
  \bibfield  {author} {\bibinfo {author} {\bibfnamefont {A.}~\bibnamefont
  {Jain}}, \bibinfo {author} {\bibfnamefont {S.~P.}\ \bibnamefont {Ong}},
  \bibinfo {author} {\bibfnamefont {G.}~\bibnamefont {Hautier}}, \bibinfo
  {author} {\bibfnamefont {W.}~\bibnamefont {Chen}}, \bibinfo {author}
  {\bibfnamefont {W.~D.}\ \bibnamefont {Richards}}, \bibinfo {author}
  {\bibfnamefont {S.}~\bibnamefont {Dacek}}, \bibinfo {author} {\bibfnamefont
  {S.}~\bibnamefont {Cholia}}, \bibinfo {author} {\bibfnamefont
  {D.}~\bibnamefont {Gunter}}, \bibinfo {author} {\bibfnamefont
  {D.}~\bibnamefont {Skinner}}, \bibinfo {author} {\bibfnamefont
  {G.}~\bibnamefont {Ceder}},\ and\ \bibinfo {author} {\bibfnamefont {K.~a.}\
  \bibnamefont {Persson}},\ }\href {https://doi.org/10.1063/1.4812323}
  {\bibfield  {journal} {\bibinfo  {journal} {APL Materials}\ }\textbf
  {\bibinfo {volume} {1}},\ \bibinfo {pages} {011002} (\bibinfo {year}
  {2013})}\BibitemShut {NoStop}%
\bibitem [{\citenamefont {A.R.~Moss}(1960)}]{arc_melting}%
  \BibitemOpen
  \bibfield  {author} {\bibinfo {author} {\bibfnamefont {D.~R.}\ \bibnamefont
  {A.R.~Moss}},\ }\href
  {https://doi.org/https://doi.org/10.1016/0022-5088(60)90024-2} {\bibfield
  {journal} {\bibinfo  {journal} {Journal of the Less Common Metals}\ }\textbf
  {\bibinfo {volume} {2}},\ \bibinfo {pages} {405} (\bibinfo {year}
  {1960})}\BibitemShut {NoStop}%
\bibitem [{\citenamefont {van~der PAUW}()}]{VanderPauw}%
  \BibitemOpen
  \bibfield  {author} {\bibinfo {author} {\bibfnamefont {L.~J.}\ \bibnamefont
  {van~der PAUW}},\ }\bibinfo {title} {A method of measuring specific
  resistivity and hall effect of discs of arbitrary shape},\ in\ \href
  {https://doi.org/10.1142/9789814503464\_0017} {\emph {\bibinfo {booktitle}
  {Semiconductor Devices: Pioneering Papers}}},\ pp.\ \bibinfo {pages}
  {174--182}\BibitemShut {NoStop}%
\bibitem [{\citenamefont {Wu}\ \emph {et~al.}(2010)\citenamefont {Wu},
  \citenamefont {Huang}, \citenamefont {Han}, \citenamefont {Gao},
  \citenamefont {Peng}, \citenamefont {Liu}, \citenamefont {Wang},
  \citenamefont {Cui},\ and\ \citenamefont {Zou}}]{Wu_VanderPauw}%
  \BibitemOpen
  \bibfield  {author} {\bibinfo {author} {\bibfnamefont {B.}~\bibnamefont
  {Wu}}, \bibinfo {author} {\bibfnamefont {X.}~\bibnamefont {Huang}}, \bibinfo
  {author} {\bibfnamefont {Y.}~\bibnamefont {Han}}, \bibinfo {author}
  {\bibfnamefont {C.}~\bibnamefont {Gao}}, \bibinfo {author} {\bibfnamefont
  {G.}~\bibnamefont {Peng}}, \bibinfo {author} {\bibfnamefont {C.}~\bibnamefont
  {Liu}}, \bibinfo {author} {\bibfnamefont {Y.}~\bibnamefont {Wang}}, \bibinfo
  {author} {\bibfnamefont {X.}~\bibnamefont {Cui}},\ and\ \bibinfo {author}
  {\bibfnamefont {G.}~\bibnamefont {Zou}},\ }\href
  {https://doi.org/10.1063/1.3374466} {\bibfield  {journal} {\bibinfo
  {journal} {Journal of Applied Physics}\ }\textbf {\bibinfo {volume} {107}},\
  \bibinfo {pages} {104903} (\bibinfo {year} {2010})}\BibitemShut {NoStop}%
\bibitem [{\citenamefont {Jayaraman}(1983)}]{jayaraman1983diamond}%
  \BibitemOpen
  \bibfield  {author} {\bibinfo {author} {\bibfnamefont {A.}~\bibnamefont
  {Jayaraman}},\ }\href@noop {} {\bibfield  {journal} {\bibinfo  {journal}
  {Reviews of Modern Physics}\ }\textbf {\bibinfo {volume} {55}},\ \bibinfo
  {pages} {65} (\bibinfo {year} {1983})}\BibitemShut {NoStop}%
\bibitem [{\citenamefont {Chijioke}\ \emph {et~al.}(2005)\citenamefont
  {Chijioke}, \citenamefont {Nellis}, \citenamefont {Soldatov},\ and\
  \citenamefont {Silvera}}]{chijioke_ruby_2005}%
  \BibitemOpen
  \bibfield  {author} {\bibinfo {author} {\bibfnamefont {A.~D.}\ \bibnamefont
  {Chijioke}}, \bibinfo {author} {\bibfnamefont {W.~J.}\ \bibnamefont
  {Nellis}}, \bibinfo {author} {\bibfnamefont {A.}~\bibnamefont {Soldatov}},\
  and\ \bibinfo {author} {\bibfnamefont {I.~F.}\ \bibnamefont {Silvera}},\
  }\href {https://doi.org/10.1063/1.2135877} {\bibfield  {journal} {\bibinfo
  {journal} {Journal of Applied Physics}\ }\textbf {\bibinfo {volume} {98}},\
  \bibinfo {pages} {114905} (\bibinfo {year} {2005})}\BibitemShut {NoStop}%
\bibitem [{\citenamefont {Akahama}\ and\ \citenamefont
  {Kawamura}(2006)}]{Akahama2006}%
  \BibitemOpen
  \bibfield  {author} {\bibinfo {author} {\bibfnamefont {Y.}~\bibnamefont
  {Akahama}}\ and\ \bibinfo {author} {\bibfnamefont {H.}~\bibnamefont
  {Kawamura}},\ }\href {https://doi.org/10.1063/1.2335683} {\bibfield
  {journal} {\bibinfo  {journal} {Journal of Applied Physics}\ }\textbf
  {\bibinfo {volume} {100}},\ \bibinfo {pages} {043516} (\bibinfo {year}
  {2006})}\BibitemShut {NoStop}%
\bibitem [{\citenamefont {Toby}\ and\ \citenamefont
  {Von~Dreele}(2013)}]{toby_gsas-ii_2013}%
  \BibitemOpen
  \bibfield  {author} {\bibinfo {author} {\bibfnamefont {B.~H.}\ \bibnamefont
  {Toby}}\ and\ \bibinfo {author} {\bibfnamefont {R.~B.}\ \bibnamefont
  {Von~Dreele}},\ }\href {http://scripts.iucr.org/cgi-bin/paper?aj5212}
  {\bibfield  {journal} {\bibinfo  {journal} {Journal of Applied
  Crystallography}\ }\textbf {\bibinfo {volume} {46}},\ \bibinfo {pages} {544}
  (\bibinfo {year} {2013})}\BibitemShut {NoStop}%
\bibitem [{\citenamefont {Zhou}\ \emph {et~al.}(2023)\citenamefont {Zhou},
  \citenamefont {Li}, \citenamefont {Shen}, \citenamefont {Feng}, \citenamefont
  {Xu}, \citenamefont {Guo}, \citenamefont {He}, \citenamefont {Qian},
  \citenamefont {Zhu},\ and\ \citenamefont {Xu}}]{multisteptransition}%
  \BibitemOpen
  \bibfield  {author} {\bibinfo {author} {\bibfnamefont {W.}~\bibnamefont
  {Zhou}}, \bibinfo {author} {\bibfnamefont {B.}~\bibnamefont {Li}}, \bibinfo
  {author} {\bibfnamefont {Y.}~\bibnamefont {Shen}}, \bibinfo {author}
  {\bibfnamefont {J.~J.}\ \bibnamefont {Feng}}, \bibinfo {author}
  {\bibfnamefont {C.~Q.}\ \bibnamefont {Xu}}, \bibinfo {author} {\bibfnamefont
  {H.~T.}\ \bibnamefont {Guo}}, \bibinfo {author} {\bibfnamefont
  {Z.}~\bibnamefont {He}}, \bibinfo {author} {\bibfnamefont {B.}~\bibnamefont
  {Qian}}, \bibinfo {author} {\bibfnamefont {Z.}~\bibnamefont {Zhu}},\ and\
  \bibinfo {author} {\bibfnamefont {X.}~\bibnamefont {Xu}},\ }\href
  {https://doi.org/10.1103/PhysRevB.108.184504} {\bibfield  {journal} {\bibinfo
   {journal} {Phys. Rev. B}\ }\textbf {\bibinfo {volume} {108}},\ \bibinfo
  {pages} {184504} (\bibinfo {year} {2023})}\BibitemShut {NoStop}%
\bibitem [{\citenamefont {{Menegotto Costa}}\ \emph {et~al.}(2013)\citenamefont
  {{Menegotto Costa}}, \citenamefont {Dias}, \citenamefont {Pureur},\ and\
  \citenamefont {Obradors}}]{MENEGOTTOCOSTA2013202}%
  \BibitemOpen
  \bibfield  {author} {\bibinfo {author} {\bibfnamefont {R.}~\bibnamefont
  {{Menegotto Costa}}}, \bibinfo {author} {\bibfnamefont {F.}~\bibnamefont
  {Dias}}, \bibinfo {author} {\bibfnamefont {P.}~\bibnamefont {Pureur}},\ and\
  \bibinfo {author} {\bibfnamefont {X.}~\bibnamefont {Obradors}},\ }\href
  {https://doi.org/https://doi.org/10.1016/j.physc.2013.09.015} {\bibfield
  {journal} {\bibinfo  {journal} {Physica C: Superconductivity}\ }\textbf
  {\bibinfo {volume} {495}},\ \bibinfo {pages} {202} (\bibinfo {year}
  {2013})}\BibitemShut {NoStop}%
\bibitem [{\citenamefont {Pei}\ \emph {et~al.}(2023)\citenamefont {Pei},
  \citenamefont {Zhang}, \citenamefont {Wang}, \citenamefont {Zhao},
  \citenamefont {Gao}, \citenamefont {Gong}, \citenamefont {Tian},
  \citenamefont {Luo}, \citenamefont {Li}, \citenamefont {Yang}, \citenamefont
  {Lu}, \citenamefont {Lei}, \citenamefont {Liu},\ and\ \citenamefont
  {Qi}}]{pei-etal-2023-supplement}%
  \BibitemOpen
  \bibfield  {author} {\bibinfo {author} {\bibfnamefont {C.}~\bibnamefont
  {Pei}}, \bibinfo {author} {\bibfnamefont {J.}~\bibnamefont {Zhang}}, \bibinfo
  {author} {\bibfnamefont {Q.}~\bibnamefont {Wang}}, \bibinfo {author}
  {\bibfnamefont {Y.}~\bibnamefont {Zhao}}, \bibinfo {author} {\bibfnamefont
  {L.}~\bibnamefont {Gao}}, \bibinfo {author} {\bibfnamefont {C.}~\bibnamefont
  {Gong}}, \bibinfo {author} {\bibfnamefont {S.}~\bibnamefont {Tian}}, \bibinfo
  {author} {\bibfnamefont {R.}~\bibnamefont {Luo}}, \bibinfo {author}
  {\bibfnamefont {M.}~\bibnamefont {Li}}, \bibinfo {author} {\bibfnamefont
  {W.}~\bibnamefont {Yang}}, \bibinfo {author} {\bibfnamefont {Z.~Y.}\
  \bibnamefont {Lu}}, \bibinfo {author} {\bibfnamefont {H.}~\bibnamefont
  {Lei}}, \bibinfo {author} {\bibfnamefont {K.}~\bibnamefont {Liu}},\ and\
  \bibinfo {author} {\bibfnamefont {Y.}~\bibnamefont {Qi}},\ }\href
  {https://doi.org/10.1093/nsr/nwad034} {\bibfield  {journal} {\bibinfo
  {journal} {Natl Sci Rev}\ }\textbf {\bibinfo {volume} {10}},\ \bibinfo
  {pages} {nwad034} (\bibinfo {year} {2023})}\BibitemShut {NoStop}%
\bibitem [{\citenamefont {Quan}()}]{quan_private_comm}%
  \BibitemOpen
  \bibfield  {author} {\bibinfo {author} {\bibfnamefont {Y.}~\bibnamefont
  {Quan}},\ }\href@noop {} {\bibinfo {title} {Private
  communication}}\BibitemShut {NoStop}%
\bibitem [{\citenamefont {Stauffer}\ and\ \citenamefont
  {Aharony}(1992)}]{Stauffer_Aharony_1992}%
  \BibitemOpen
  \bibfield  {author} {\bibinfo {author} {\bibfnamefont {D.}~\bibnamefont
  {Stauffer}}\ and\ \bibinfo {author} {\bibfnamefont {A.}~\bibnamefont
  {Aharony}},\ }\href {https://doi.org/10.1201/9781315274386} {\emph {\bibinfo
  {title} {Introduction To Percolation Theory}}},\ \bibinfo {edition} {2nd}\
  ed.\ (\bibinfo  {publisher} {Taylor \& Francis},\ \bibinfo {year}
  {1992})\BibitemShut {NoStop}%
\bibitem [{\citenamefont {Lobb}\ \emph {et~al.}(1978)\citenamefont {Lobb},
  \citenamefont {Tinkham},\ and\ \citenamefont {Skocpol}}]{LOBB19781273}%
  \BibitemOpen
  \bibfield  {author} {\bibinfo {author} {\bibfnamefont {C.}~\bibnamefont
  {Lobb}}, \bibinfo {author} {\bibfnamefont {M.}~\bibnamefont {Tinkham}},\ and\
  \bibinfo {author} {\bibfnamefont {W.}~\bibnamefont {Skocpol}},\ }\href
  {https://doi.org/https://doi.org/10.1016/0038-1098(78)91550-8} {\bibfield
  {journal} {\bibinfo  {journal} {Solid State Communications}\ }\textbf
  {\bibinfo {volume} {27}},\ \bibinfo {pages} {1273} (\bibinfo {year}
  {1978})}\BibitemShut {NoStop}%
\bibitem [{\citenamefont {Lobb}(1980)}]{osti_6602588}%
  \BibitemOpen
  \bibfield  {author} {\bibinfo {author} {\bibfnamefont {C.~J.}\ \bibnamefont
  {Lobb}}\ }\href {https://www.osti.gov/biblio/6602588} {} (\bibinfo {year}
  {1980})\BibitemShut {NoStop}%
\bibitem [{\citenamefont {Krivoruchko}\ and\ \citenamefont
  {Tarenkov}(2019)}]{Krivoruchko2019}%
  \BibitemOpen
  \bibfield  {author} {\bibinfo {author} {\bibfnamefont {V.~N.}\ \bibnamefont
  {Krivoruchko}}\ and\ \bibinfo {author} {\bibfnamefont {V.~Y.}\ \bibnamefont
  {Tarenkov}},\ }\href {https://doi.org/10.1063/1.5097355} {\bibfield
  {journal} {\bibinfo  {journal} {Low Temperature Physics}\ }\textbf {\bibinfo
  {volume} {45}},\ \bibinfo {pages} {476} (\bibinfo {year} {2019})}\BibitemShut
  {NoStop}%
\bibitem [{\citenamefont {Sternfeld}\ \emph {et~al.}(2005)\citenamefont
  {Sternfeld}, \citenamefont {Shelukhin}, \citenamefont {Tsukernik},
  \citenamefont {Karpovski}, \citenamefont {Gerber},\ and\ \citenamefont
  {Palevski}}]{Sternfeld2005}%
  \BibitemOpen
  \bibfield  {author} {\bibinfo {author} {\bibfnamefont {I.}~\bibnamefont
  {Sternfeld}}, \bibinfo {author} {\bibfnamefont {V.}~\bibnamefont
  {Shelukhin}}, \bibinfo {author} {\bibfnamefont {A.}~\bibnamefont
  {Tsukernik}}, \bibinfo {author} {\bibfnamefont {M.}~\bibnamefont
  {Karpovski}}, \bibinfo {author} {\bibfnamefont {A.}~\bibnamefont {Gerber}},\
  and\ \bibinfo {author} {\bibfnamefont {A.}~\bibnamefont {Palevski}},\ }\href
  {https://doi.org/10.1103/PhysRevB.71.064515} {\bibfield  {journal} {\bibinfo
  {journal} {Phys. Rev. B}\ }\textbf {\bibinfo {volume} {71}},\ \bibinfo
  {pages} {064515} (\bibinfo {year} {2005})}\BibitemShut {NoStop}%
\bibitem [{\citenamefont {Deutscher}\ and\ \citenamefont
  {de~Gennes}(1969)}]{Deutscher_Gennes_1969}%
  \BibitemOpen
  \bibfield  {author} {\bibinfo {author} {\bibfnamefont {G.}~\bibnamefont
  {Deutscher}}\ and\ \bibinfo {author} {\bibfnamefont {P.~G.}\ \bibnamefont
  {de~Gennes}},\ }in\ \href@noop {} {\emph {\bibinfo {booktitle}
  {Superconductivity}}}\ (\bibinfo  {publisher} {Routledge},\ \bibinfo {year}
  {1969})\ \bibinfo {edition} {1st}\ ed.,\ p.~\bibinfo {pages} {30}\BibitemShut
  {NoStop}%
\bibitem [{\citenamefont {Seleznyov}\ \emph {et~al.}(2024)\citenamefont
  {Seleznyov}, \citenamefont {Yagovtsev}, \citenamefont {Pugach},\ and\
  \citenamefont {Tao}}]{SELEZNYOV2024171645}%
  \BibitemOpen
  \bibfield  {author} {\bibinfo {author} {\bibfnamefont {D.}~\bibnamefont
  {Seleznyov}}, \bibinfo {author} {\bibfnamefont {V.}~\bibnamefont
  {Yagovtsev}}, \bibinfo {author} {\bibfnamefont {N.}~\bibnamefont {Pugach}},\
  and\ \bibinfo {author} {\bibfnamefont {L.}~\bibnamefont {Tao}},\ }\href
  {https://doi.org/https://doi.org/10.1016/j.jmmm.2023.171645} {\bibfield
  {journal} {\bibinfo  {journal} {Journal of Magnetism and Magnetic Materials}\
  }\textbf {\bibinfo {volume} {595}},\ \bibinfo {pages} {171645} (\bibinfo
  {year} {2024})}\BibitemShut {NoStop}%
\bibitem [{\citenamefont {Hamlin}\ \emph {et~al.}(2024)\citenamefont {Hamlin},
  \citenamefont {Sinha}, \citenamefont {Lim},\ and\ \citenamefont
  {Li}}]{hamlin_2024_13359794}%
  \BibitemOpen
  \bibfield  {author} {\bibinfo {author} {\bibfnamefont {J.}~\bibnamefont
  {Hamlin}}, \bibinfo {author} {\bibfnamefont {S.}~\bibnamefont {Sinha}},
  \bibinfo {author} {\bibfnamefont {J.}~\bibnamefont {Lim}},\ and\ \bibinfo
  {author} {\bibfnamefont {Z.}~\bibnamefont {Li}},\ }\href
  {https://doi.org/10.5281/zenodo.13359794} {\bibinfo {title}
  {jhamlin-ufl/remob2: v0.9.2}} (\bibinfo {year} {2024})\BibitemShut {NoStop}%
\end{thebibliography}

%

\end{document}


\begin{CJK*}{UTF8}{gbsn}

	\title{{Supplemental Material: Superconductivity in pressurized \ReMoB}}

	\author{S. Sinha}
	\affiliation{Department of Physics, University of Florida, Gainesville, Florida 32611, USA}
	\author{J. Lim}
	\affiliation{Department of Physics, University of Florida, Gainesville, Florida 32611, USA}
	\author{Z. Li}
	\affiliation{Department of Physics, University of Florida, Gainesville, Florida 32611, USA}
	\author{J. S. Kim}
	\affiliation{Department of Physics, University of Florida, Gainesville, Florida 32611, USA}
	\author{A. C. Hire}
	\affiliation{Department of Materials Science and  Engineering, University of Florida, Gainesville, Florida 32611, USA}
	\affiliation{Quantum Theory Project, University of Florida, Gainesville, Florida 32611, USA}
	\author{P. M. Dee}
	\affiliation{Department of Physics, University of Florida, Gainesville, Florida 32611, USA}
	\affiliation{Department of Materials Science and  Engineering, University of Florida, Gainesville, Florida 32611, USA}
	\author{R. S. Kumar}
	\affiliation{Department of Physics, University of Illinois Chicago, Chicago, Illinois 60607, USA}
	\author{D.~Popov}
	\affiliation{HPCAT, X-ray Science Division, Argonne National Laboratory, Argonne, Illinois 60439, USA}
	\author{R. J. Hemley}
	\affiliation{Department of Physics, Chemistry, and Earth and Environmental Sciences, University of Illinois Chicago, Chicago, Illinois 60607, USA}
	\author{R. G. Hennig}
	\affiliation{Department of Materials Science and  Engineering, University of Florida, Gainesville, Florida 32611, USA}
	\affiliation{Quantum Theory Project, University of Florida, Gainesville, Florida 32611, USA}
	\author{P. J. Hirschfeld}
	\affiliation{Department of Physics, University of Florida, Gainesville, Florida 32611, USA}
	\author{G. R. Stewart}
	\affiliation{Department of Physics, University of Florida, Gainesville, Florida 32611, USA}
	\author{J. J. Hamlin}
	\affiliation{Department of Physics, University of Florida, Gainesville, Florida 32611, USA}
	\date{\today}
	\maketitle
\end{CJK*}

\section{Raman Pressure Determination}
Raman spectrum was recorded after shining red laser on to the sample inside diamond anvil cell using a microscope with 20X and 40X magnification.
Spectra from both 20X and 40X were plotted along with their derivatives with respect to Raman shift.
The Raman Shift at right end of the dip in derivative was fit to the equation in order to calculate pressure ~\cite{Akahama2006,diamond_raman_spectroscopy}.
These measurements were performed at room temperature and pressure at 10 K was estimated using these values.
\begin{figure}[h!]
	\centering
	\begin{subfigure}[b]{0.35\textwidth}
		\includegraphics[width=\columnwidth]{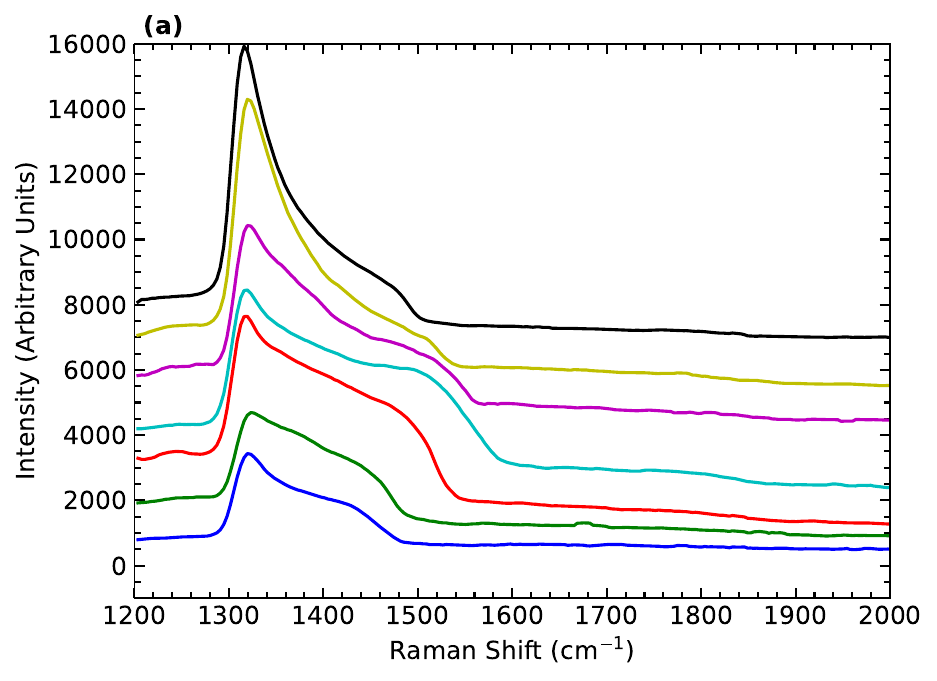}

	\end{subfigure}
	\begin{subfigure}[b]{0.35\textwidth}
		\includegraphics[width=\columnwidth]{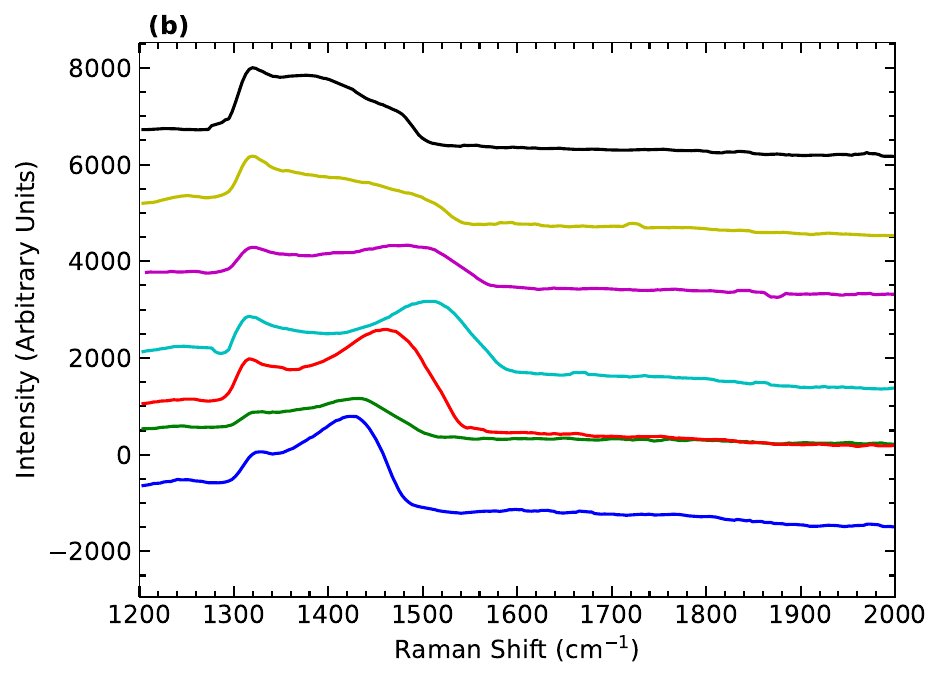}
	\end{subfigure}
	\begin{subfigure}[b]{0.35\textwidth}
		\includegraphics[width=\columnwidth]{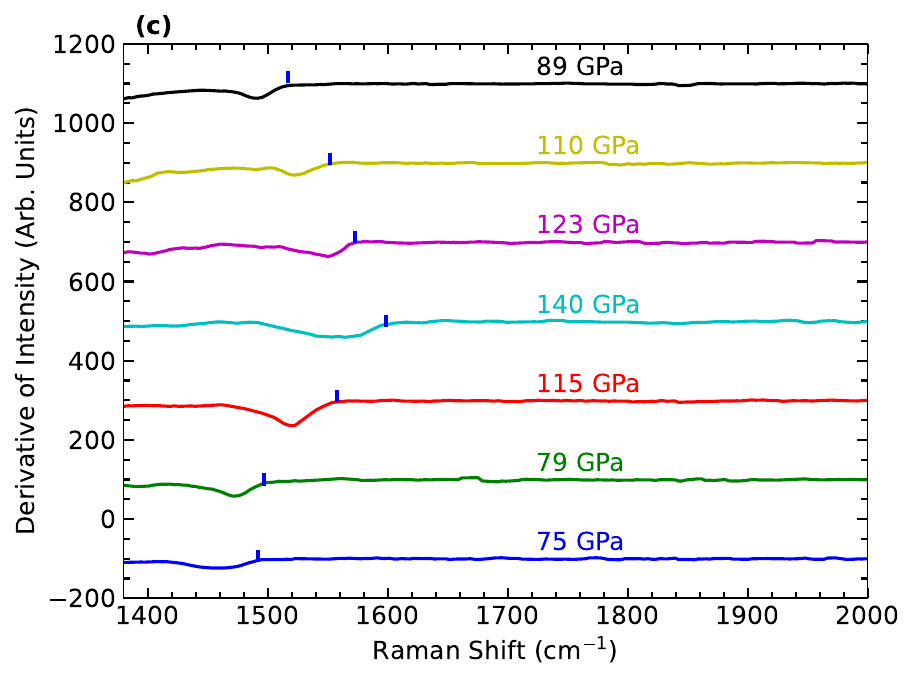}
	\end{subfigure}
	\begin{subfigure}[b]{0.35\textwidth}
		\includegraphics[width=\columnwidth]{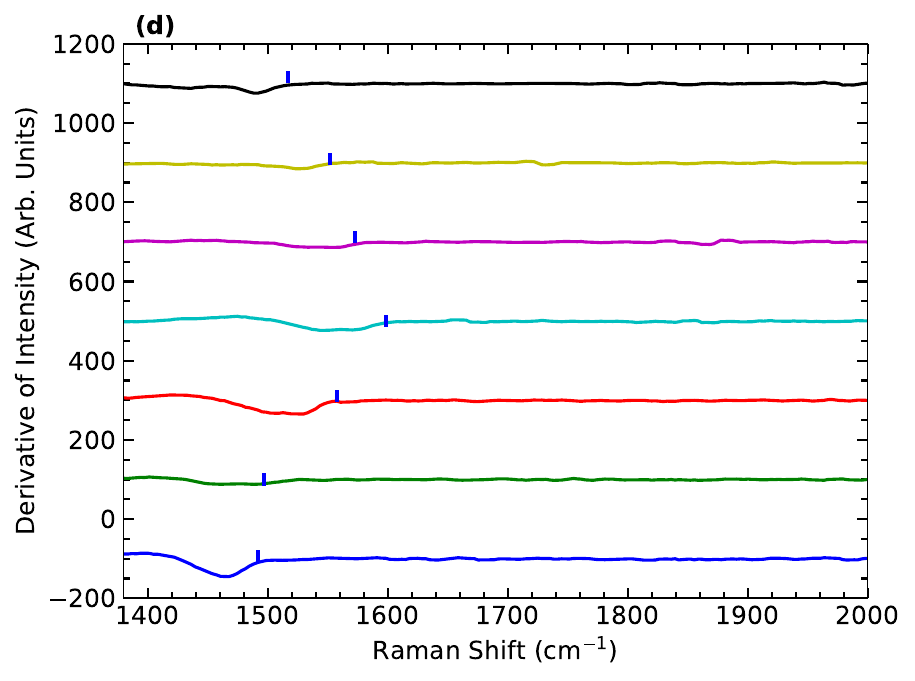}
	\end{subfigure}
	\caption{(a) Raman spectrum for pressure measurement using $20 \times$ magnification vs Raman shift for \ReMoB\ (b) Raman spectrum for pressure measurement using $40 \times$ magnification for \ReMoB\. (c) Derivative of Raman peak intensity with $20 \times$ magnification. (d) Derivative of Raman peak intensity with $40 \times$ magnification. Raman shift at ambient pressure($\nu_0$) was \SI{1332}{\per\centi\meter}. Due to stronger Raman signal in $20 \times$ magnification, it was used for pressure estimation.
		Blue tic marks in (c) and (d) indicate the Raman shift used to estimate the pressure.
	}
	\label{fig:XRD_S}
\end{figure}

\clearpage
\section{Resistance Measurements}
\begin{figure}[h!]
	\centering
	\begin{subfigure}[b]{0.48\textwidth}
		\caption{}
		\includegraphics[width=\columnwidth]{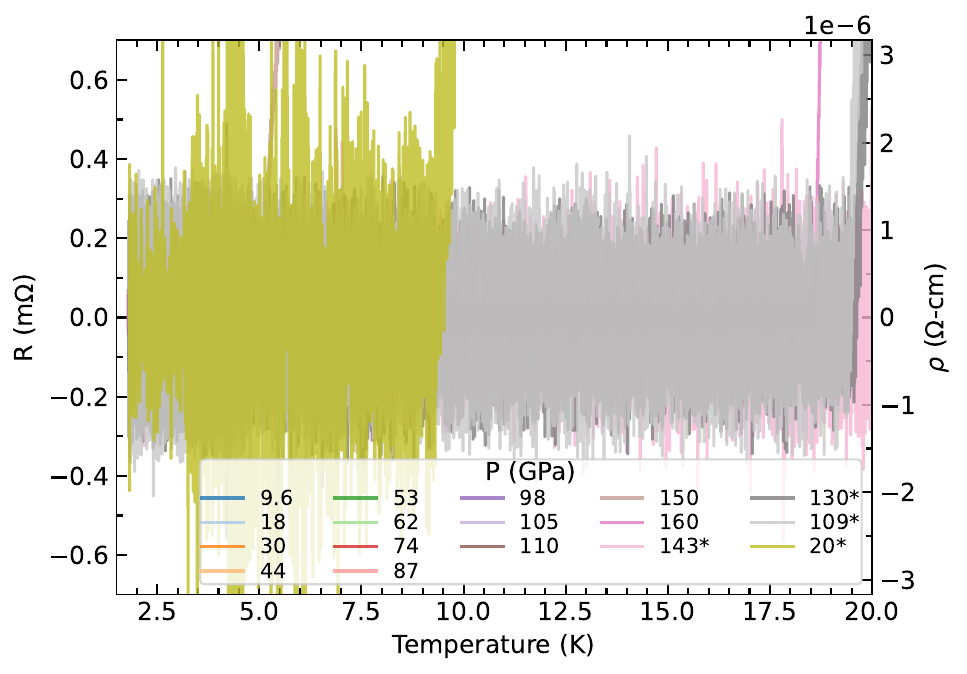}

	\end{subfigure}
	\begin{subfigure}[b]{0.46\textwidth}
		\caption{}
		\includegraphics[width=\columnwidth]{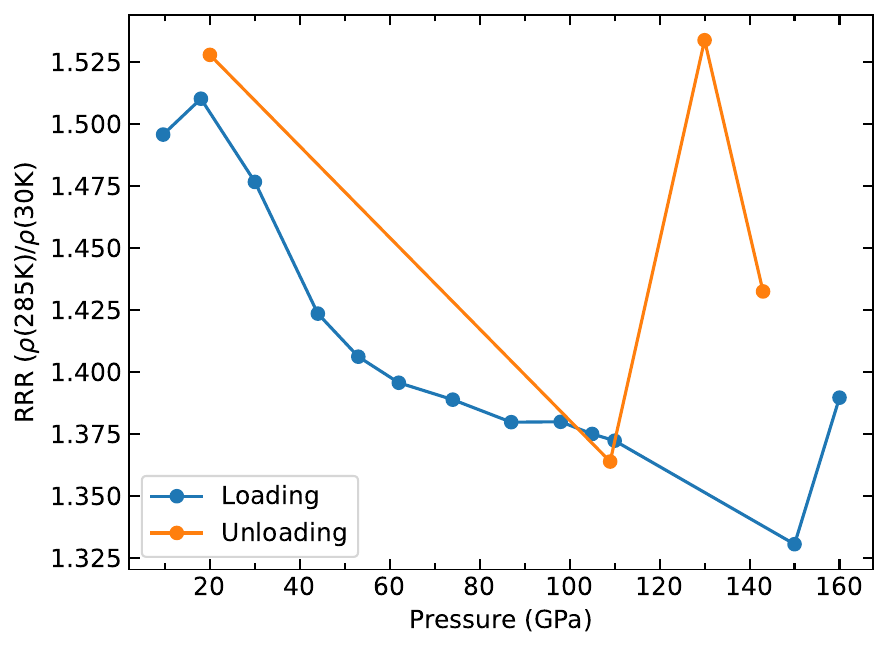}
	\end{subfigure}
	\caption{(a) Noise in resistance (R)/resistivity ($\rho$) around 0 after the superconducting transition is complete. (b) Residual Resistance Ratio (RRR) calculated using (resistance at 285 K/ resistance at 30 K).}
	\label{fig:Supp_res}
\end{figure}

Figure~\ref{fig:Supp_res}a highlights the resistance below $T_c$.
The residual resistivity ratio (RRR) depends only weakly on pressure (Fig.~\ref{fig:Supp_res}b).
\begin{figure}[h]
	\centering
	\includegraphics[width=0.9\linewidth]{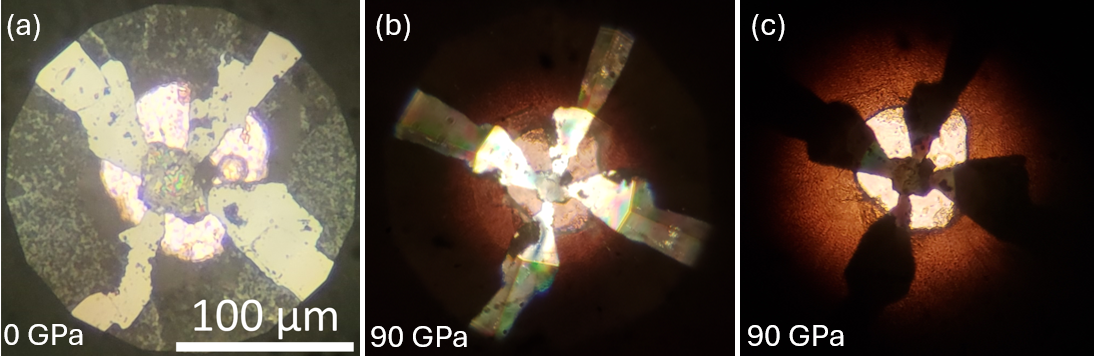}
	\caption{Sample of \ReMoB\ loaded in the diamond cell. (a) Sample is at near ambient pressure.  Four electrodes are visible as is the ruby (to right of sample).  The pressure medium is solid steatite. (b) and (c) Sample under high pressure of \SI{90}{GPa}. In case of (c), light is incident only from below.}
	\label{fig:sample}
\end{figure}

\newpage

%